\documentclass[11pt,twoside]{article}
\usepackage{newpasp,natbib,natbibspacing,graphicx,amssymb}
\usepackage[english]{babel}

\begin{document}
\title{The Unidentified InfraRed Features after ISO}
\author{E. Peeters, L.J. Allamandola, D.M. Hudgins}
\affil{NASA-Ames Research Center, Space Science Division, MS: 245-6,
  Moffett Field CA 94035-1000, U.S.A.}
\author{S. Hony}
\affil{RSSD-ESA/ESTEC, PO Box 299, 2200 AG Noordwijk, The Netherlands}
\author{A.G.G.M. Tielens}
\affil{Kapteyn Institute \& SRON National Institute for Space Research,
  P.O. Box 800, 9700 AV Groningen, The Netherlands}

\begin{abstract}
  The Infrared Space Observatory (ISO) has provided the first complete
  mid-IR spectra for a wide range of objects.  Almost all of these
  spectra are dominated by the well-known infrared emission features
  at 3.3, 6.2, 7.7, and 11.2~$\mu$m, the so-called Unidentified
  Infra-Red (UIR) features.  Besides the major features, there is an
  array of minor features and broad plateaux stretching from 3 to
  20~$\mu$m which reveal subtle details of conditions in the emission
  zones and properties of the carriers. Generally attributed to the
  vibrational relaxation of UV-pumped Polycyclic Aromatic Hydrocarbon
  molecules (PAHs) containing some 50--100 C-atoms, these UIR spectra
  are a treasure trove of information.
  
  The ISO spectra have, for the first time, allowed a systematic
  analysis of the spectral characteristics of the UIR features in a
  wide variety of environments.  The peak positions, profiles, and
  relative strengths of the major features vary from source to source
  and spatially within sources.  Variations in the 6.2 and 7.7~$\mu$m
  features are particularly noteworthy; the former peaks between
  6.25--6.3~$\mu$m in many PNe, but falls at 6.2~$\mu$m in RNe and
  \ion{H}{ii}~regions.  Similarly, the broad 7.7~$\mu$m feature
  consists of major components at 7.6 and ``7.8''~$\mu$m whose
  relative strengths vary.  The 7.6~$\mu$m component dominates in
  \ion{H}{ii}~regions and reflection nebulae, while the ``7.8''~$\mu$m
  feature takes over in most PNe. These specific profiles are not
  unique to certain object types but can occur within each individual
  source. While the 3.3 and 11.2~$\mu$m also show variations in peak
  position and profile, these are much less pronounced. In addition,
  the 3.3~$\mu$m feature intensity correlates quite well with that of
  the 11.2~$\mu$m feature, a correlation which does not extend to the
  CH modes between 12 and 14 $\mu$m. Also, variations occurs in the
  relative strengths of the CC modes in the 6--9~$\mu$m range relative
  to the CH modes 3.3 and 11.2~$\mu$m.  Such behavior requires that
  the UIR features are carried by a family of related compounds whose
  detailed physical and/or chemical characteristics vary in response
  to local physical conditions.  Here, we review ISO and recent
  ground-based observations and assess some of their implications.
\end{abstract}

\section{Introduction}
\label{EP:intro}
The infrared spectra of a wide variety of objects associated with dust
and gas -- including \ion{H}{ii}~regions, post-AGB stars, PNe, young
stellar objects (YSOs), the diffuse ISM and galaxies -- are dominated
by emission features at 3.3, 6.2, 7.7, 8.6, 11.2 and 12.7~$\mu$m,
\citep[cf.][]{Gillett:73, Geballe:85, Cohen:co:86, isoparijs}. Often,
these features are accompanied by broad underlying emission plateaux.
Since the carriers of these features remained a mystery for almost a
decade, the initial name for these features, the Unidentified InfraRed
(UIR) emission features, is still in use. However, these is also an
unfortunate proliferation of other names (Table~\ref{EP:names}). Here
we use the term UIRs to refer to these features.

In the early '80s, it was recognized that the UIR features coincide with
the vibrational modes characteristic of aromatic materials
\citep{Duley:aromatic:81}. Since then, many different carriers have
been proposed such as Hydrogenated Amorphous Carbon \citep[HAC,
e.g.][]{Duley:83, Borghesi:amcarbon:87}, Quenched Carbon Composites
\citep[QCC, e.g.][]{Sakata:QCC:84}, Polycyclic Aromatic Hydrocarbons
\citep[PAHs, e.g.][]{Allamandola:rev:89, Puget:revpah:89}, Coal
\citep[e.g.][]{Papoular:coalmodel:89}, nanodiamonds
\citep{Jones:nanodiamonds:00}, Rydberg matter
\citep{Holmlid:rydbergmatter:00} and most recently Locally Aromatic
Polycyclic Hydrocarbons \citep{Petrie:03}.

While the vibrational spectrum of (almost) any exclusively aromatic
material can provide a global fit to the observed UIR features, they
are generally attributed to {\it small} aromatic hydrocarbon
species. The main argument for this is the detection of the UIR
features in reflection nebulae far from illuminating stars and the
independence of the color temperature with the distance from the
star. Taken together, these observations indicate that the carriers of
the UIR features can be excited to very high ``temperatures'' in such
cold environments upon absorption of a {\it single} FUV photon
\citep{Sellgren:84}. This requires the emitting species to be very
small, containing a finite number of oscillators. Classical
0.1 $\mu$m size grains are too large and do not become hot enough
to emit at mid-infrared wavelengths in these environments. Hence the
UIR features are attributed to species of the order of 10\AA\,
\citep{Sellgren:84}. This corresponds to molecules containing $\sim$
50--100 carbon atoms \citep[e.g.][]{Tielens:93, Tielens:parijs:99}.
This aspect of the UIR feature phenomenon has now been reinforced by
ISO and IRTS observations which reveal that the mid-infrared spectra
of the (diffuse) ISM is dominated by these IR emission features
\citep[see Onaka, these proceedings;][]{Mattila:96,Onaka:dism:96,
Boulanger:98, Boulanger:leshouches:99, Boulanger:00, Onaka:00}.  It
should be emphasized that, at the smallest scales, all of the proposed
grain carriers (HAC, QCC, coal, ...) share an aromatic hydrocarbon
structure with PAH molecules.  However, some of these carriers may
vary in detail from PAH molecules (adopting their chemical
definition), even at sizes required by the energetics, by including
aliphatic hydrocarbon chains, by containing substitution of other
elements such as N or O for a C atom in a PAH, or by having D or
molecular sidegroups substituting for an H atom in a PAH.  However,
limits on the presence of other functional groups are very strict
\citep{Tielens:parijs:99}. Methyl groups (-CH$_3$) are the most
abundant at 0.02 of aromatic H (or equivalently $\sim$0.01 relative to
aromatic C). Other substitutions such as OH and NH$_2$ are less
abundant than 0.01 relative to aromatic H. There is some spectroscopic
evidence for the presence of oxygen bonded in the form of carbonyl
groups (-C=O; much less then 0.01 relative to aromatic C). All other
species are at a very low level.

In short, the excitation of such small species should be discussed in
molecular physics terms.  An absorption of a FUV photon by a small
aromatic hydrocarbon induces a transition to an upper electronic
state. The excited molecule then makes rapid isoenergetic transitions
to highly vibrationally excited levels of the ground electronic
state. Subsequently, this highly vibrationally excited molecule
relaxes, mainly by IR emission in the CC and CH fundamental
vibrational modes \citep[][see Li, these
proceedings]{Allamandola:rev:89}. In this way, small aromatic
hydrocarbons leave their signature in the form of the UIR features.

\begin{table}[!t]
\caption{Possible names for the IR emission band family found in the
literature to date.}
\label{EP:names}
\begin{center}
\begin{tabular}{ll}
& \\[-5pt]
\hline
& \\[-5pt]
UIRs         & Un-identified InfraRed emission features\\
IEFs / IEBs  & Infrared Emission Features/Bands\\
AEFs / AFE   & Aromatic Emission Features / Aromatic Feature Emission\\
AIBs      & Aromatic Infrared (emission) Bands \\
PAHs      & Polycyclic Aromatic Hydrocarbon bands/features\\
OIRs      & Over-identified InfraRed emission features\\[5pt]
\hline
\end{tabular}
\end{center}
\end{table}

Mid-infrared spectra from many objects are dominated by the UIR
emission features. Remarkably 20-30\% of the Galactic IR radiation is
emitted in these UIR features and 10-15\% of the cosmic carbon is
locked up in the UIR carriers \citep{Snow:c:95}.

Being so widespread and abundant, they play a crucial role in several
astrophysical processes.  Aromatic units are the building blocks in
the cosmic carbon condensation route \citep{Frenklach:form:89,
Tielens:stardust:90, Tielens:carbonstardust:97, Tielens:organicmol:97,
Cherchneff:dustformwr:00}.  The UIR carriers dominate the heating and
cooling of the ambient interstellar medium (ISM) via photoelectric ejection
\citep{Verstraete:90, Bakes:photoelec:94}, infrared emission and
gas-grain collisional cooling \citep{Aannestad:79, Dwek:86}. In
addition, they influence the charge balance and in this way -- through
their influence on the equilibrium state of chemical reactions --
gas-phase abundances in interstellar (IS) clouds \citep{Lepp:88, Bakes:98}.
In particular, they might reduce the gas-phase D/H ratio
\citep{Draine:dust:03, Peeters:pads:04}.  Due to their large surface
areas, they also affect the on-going surface chemistry and so play a
significant role in IS chemistry \citep{Tielens:dust:87}. Lastly, they
are an important, chemically accessible, source of prebiotic organic
material.

Although the presence of aromatic hydrocarbon species in space is now
generally accepted, today, specifics of the emitting population
remain elusive. Progress is being made however (e.g. Hudgins \&
Allamandola, these proceedings) and strong constraints are placed on
molecular structures \citep{Allain:ionendehydro:96, Hony:oops:01},
sizes \citep{Schutte:model:93}, heteroatom substitution
\citep{Peeters:prof6:02, Bauschlicher:NPAHs:04}, charge balance
\citep{Allamandola:modelobs:99,Wagner:2000,LePage:03} and related
properties \citep{Robinson:97,Crete:99, Chan:01}.  The combination of
the astronomical observations with the laboratory and theoretical
studies increases our understanding of how the local environment
and/or the history influence the composition of this population; and
vice-versa, how the aromatic hydrocarbon molecules influence the local
physical and chemical processes. Such insight is also essential if we
want to use the UIR emission features as tools for probing the
universe.

Many studies have attempted to solve pieces of this puzzle and here we
review the progress made in the observational domain since the launch
of the Infrared Space Observatory (ISO). With its complete access to
the 2.3--197~$\mu$m, wavelength range, ISO allowed the study of the
UIR features in all their glory. Complemented with observations by the
IRTS and recent ground-based instruments, these ISO observations have
opened the doorway to the exploitation of the UIR features as probes of a
wide variety of astronomical environments -- from RNe to galaxies.

In the following, Sect. 2 highlights the omnipresent and rich UIR
spectrum. Its spectral variations and their implications are
extensively discussed in Sects. 3 and 4. The UIR emission features as
diagnostic tools are considered in Sect. 5. Finally, Sect. 6 gives a
summary and looks to the future.

\section{The UIR spectrum}
\label{EP:richsp}
\begin{figure}[t!]
\includegraphics[clip,angle=90,width=\textwidth]{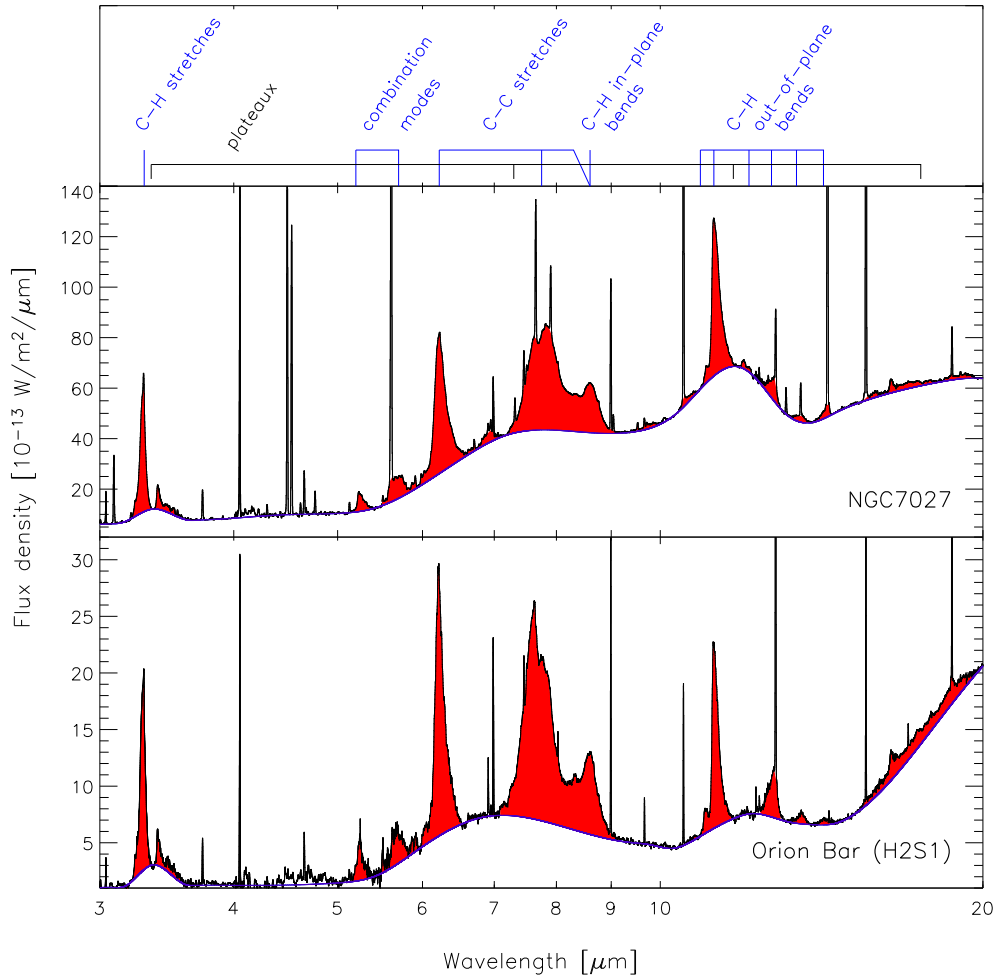}
\caption{The ISO-SWS spectra of the planetary nebula NGC~7027 and the
Photo-Dissociation region at the Orion Bar illustrate the richness and
variety of the UIR spectrum. Also indicated are the aromatic mode
identifications of the major UIR features.}
\label{EP:rich}
\end{figure}

Fig.~\ref{EP:rich} illustrates the spectral richness and some of the
variations of the UIR spectra. Clearly, many weaker features are
present apart from the major UIR features at 3.3, 6.2, 7.7, 8.6, 11.2
and 12.7~$\mu$m. Some of these features are perched on top of emission
plateaux of variable strength. These plateaux are treated in 2
ways. Either they are assumed to be part of the individual features
and fitted with Lorentzians in combination with a power-law continuum
\citep[Fig.~\ref{EP:joblin}, e.g.][]{Boulanger:lorentz:98,
Verstraete:prof:01}. Or, they are considered independently of the
individual features and then local spline continua are used
\citep[Fig.~\ref{EP:rich}, e.g.][]{Hony:oops:01,
Peeters:prof6:02}. The latter is based upon early observations
revealing a different spatial distribution for the plateaux and the
features \citep{Bregman:orion:89}.  Adopting the latter method, the
3.3~$\mu$m feature is situated on a broad plateau and is often
accompanied by features at 3.4 and/or 3.5~$\mu$m. Two weaker features
are found at 5.2 and 5.7~$\mu$m. The 6.2 and 11.2~$\mu$m features are
accompanied with satellite features at 6.0, 11.0 and sometimes at
$\sim$6.6~$\mu$m. The broad 7.7~$\mu$m complex is composed of at least
2 subcomponents at 7.6 and 7.8~$\mu$m. Some sources exhibit more
substructures e.g. near 7.2 to 7.4 and 8.2~$\mu$m
\citep{Moutou:leshouches:99, Moutou:c60:99, Peeters:prof6:02}. A broad
emission plateau of variable strength is present underneath the 6.2,
7.7 and 8.6~$\mu$m features. The onset of this emission plateau seems
to be variable and falls longwards of $\sim$ 6~$\mu$m while it extends
to $\sim$ 9~$\mu$m. The out-of-plane CH bending modes are present at
10.8, 11.0, 11.2, 12.0, 12.7, 13.2 and 14.5~$\mu$m and also sit on a
broad emission plateau. At even longer wavelengths, a new feature has
been reported at 16.4~$\mu$m \citep{Moutou:16.4:00,
VanKerckhoven:plat:00} as well as a weak emission plateau between 15
and 20~$\mu$m \citep{VanKerckhoven:plat:00}. It should be
emphasized that not all sources show all these emission features at
the same time and their peak position and relative strength
vary as discussed below.

The importance of this family of emission features cannot be over
emphasized. These emission features are now found in almost all
environments including diffuse ISM, the edges of molecular clouds,
reflection nebulae, young stellar objects, \ion{H}{ii}~regions, star forming
regions, some C-rich WR stars, post-AGB stars, planetary nebulae,
novae, normal galaxies, starburst galaxies, most Ultra-Luminous IR
galaxies (ULIRGs) and AGNs. Thus, it is no exaggeration to say that the
niche for the UIR carriers is the universe.

\section{Feature Positions and Profiles}
\label{EP:prof}
Until quite recently, most of the interstellar emission features were
considered to be more-or-less invariant in position and profile. The
3.3~$\mu$m feature is the exception. High resolution observations by
\citet{Tokunaga:33prof:91} and \citet{Kerr:rr:99} showed slight
variations in peak position and profile. Although some minor
variations were noted, by and large the 6.2~$\mu$m feature was
considered fixed at 6.2~$\mu$m, regardless of the reported shift in
peak position by \citet{Molster:pah:96}. The "7.7"~$\mu$m feature was
generally treated similarly in spite of earlier papers showing this
feature is comprised of at least two variable components
\citep[e.g.][]{Bregman:hiivspn:89, Cohen:southerniras:89}.  It has
long been recognized that the 7.7~$\mu$m complex appears either with a
dominant 7.6~$\mu$m component or with the dominant component peaking
at 7.8--8~$\mu$m \citep{Bregman:hiivspn:89, Cohen:southerniras:89,
Molster:pah:96, Roelfsema:pahs:96, Moutou:leshouches:99,
Moutou:parijs:99, Peeters:parijs:99}. The former profile is associated
with \ion{H}{ii}~regions and the latter with planetary nebulae
\citep{Bregman:hiivspn:89, Cohen:southerniras:89}.

\subsection{Source to source variations}
\label{EP:source-source-vari}
\begin{figure}[t!]
\includegraphics[clip,angle=90,width=\textwidth]{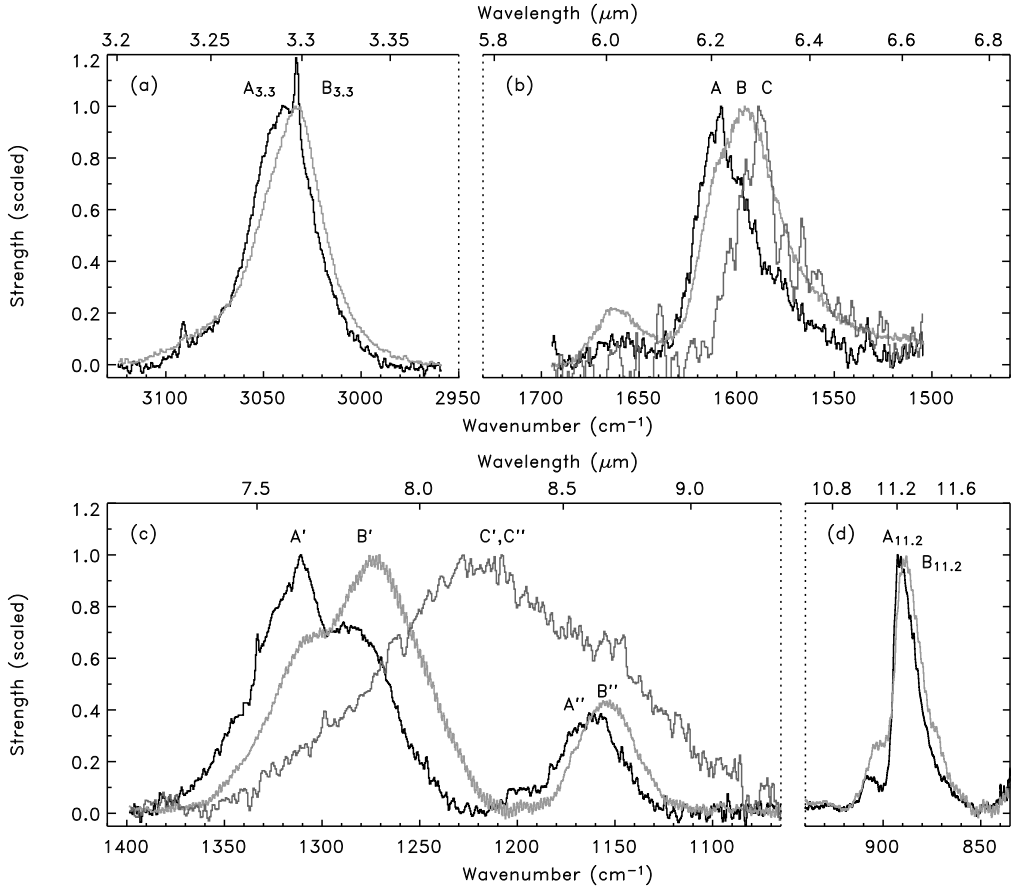}
\caption{An overview of the source to source
  variations in the position and profile of the 3--12~$\mu$m UIR
  features. For all features, class $A$ peaks at the shortest
  wavelengths (black line), and class $B$ peaks at longer wavelength.
  In the 6--9~$\mu$m region, another class ($C$) is defined peaking at
  even longer wavelength with a unique spectral appearance. For
  details see \citet{Peeters:prof6:02} and
  \citet{vanDiedenhoven:chvscc:03}.  }
\label{EP:prof_sws}
\end{figure}

Using ISO-SWS spectra, \citet{Peeters:prof6:02} and
\citet{vanDiedenhoven:chvscc:03} have carried out a detailed study of
the well-known UIR features in a large sample of reflection nebulae,
\ion{H}{ii}~regions, YSOs, evolved stars and galaxies.  These authors
found that the peak position of the UIR features clearly varies from
source to source (see Fig.~\ref{EP:prof_sws}).  In addition, distinct
differences are found in their profiles. In particular, the 6.2~$\mu$m
profile is very asymmetric with a steep blue wing and a red tail when
peaking at the shortest wavelengths (6.22 $\mu$m) but it is generally
far more symmetric when peaking at the longest wavelengths
(i.e. 6.3~$\mu$m). Nevertheless, more asymmetric profiles peaking at
6.3--6.4~$\mu$m are also reported for hydrogen deficient environments
\citep{Harrington:64:98, Chiar:wr:02}. The 7.7~$\mu$m complex is
indeed comprised of two components, one at 7.6 $\mu$m and one at
``7.8'' $\mu$m. The feature varies from a profile with a dominant to a
minor 7.6~$\mu$m component while simultaneously shifting, as a whole,
to longer wavelengths. The peak position of the ``7.8'' component is
highly variable and ranges from 7.8 up to 8~$\mu$m. Several sources
show an additional subcomponent between 7.2 and 7.4~$\mu$m \citep[see
also][]{Moutou:leshouches:99, Moutou:c60:99}. The profile of the
8.6~$\mu$m feature is clearly symmetric and similar for all
sources. However, distinct shifts in peak position are also observed
for this feature. Four sources are known to date that show a peculiar
8.6~$\mu$m feature \citep[see also][]{Roelfsema:pahs:96,
Verstraete:m17:96, Peeters:sc18434:04}.  A very limited number of
sources does not show a 7.7~$\mu$m complex nor a 8.6~$\mu$m
feature but instead exhibit a broad emission feature peaking at
8.22~$\mu$m. This is clearly distinct from the 8~$\mu$m emission
feature found in some post-AGB stars.  The 3.3~$\mu$m profile is
clearly symmetric when positioned at the shortest wavelengths and
asymmetric when peaking at slightly longer wavelengths. The former
profile has a similar peak position and FWHM as the Type 1 profile of
\citet{Tokunaga:33prof:91}, the latter has a peak position similar to
their Type 2 but is much wider. The 11.2~$\mu$m profile comprises a
steep blue rise and a red tail and only shows small variations in both
peak position and profile.

The observed pronounced contrast in the spectral variations for the CH
modes versus the CC modes is striking : the peak wavelengths of the
features attributed to CC modes (6--9 $\mu$m) vary by $\sim$25--50
cm$^{-1}$, while for the CH modes (3.3 and 11.2 $\mu$m features) the
variations are smaller, $\sim$4--6.5 cm$^{-1}$. Moreover, these
profile variations in the CC modes are directly linked with each
other, i.e. the 6.2~$\mu$m profile A is correlated with the 7.7~$\mu$m
profile A$^{\prime}$ and the 8.6~$\mu$m profile A$^{\prime\prime}$
while the 6.2~$\mu$m profile B is correlated with the 7.7~$\mu$m
profile B$^{\prime}$ and the 8.6~$\mu$m profile B$^{\prime\prime}$. In
contrast, the CH feature profile variations appear less connected to
each other or to those found for the CC modes.

As already noted by \citet{Bregman:hiivspn:89} and
\citet{Cohen:southerniras:89} for the 7.7~$\mu$m complex, the specific
profile of the CC modes (but less for the CH modes) depend directly on
the type of object.  All \ion{H}{ii}~regions, non-isolated YSOs and
reflection nebulae exhibit class A emission features (black line in
Fig.~\ref{EP:prof_sws}). Lines-of-sight towards the (diffuse) ISM also
belong to this group \citep[see Onaka, these
proceedings,][]{Mattila:96, Moutou:parijs:99,
Uchida:RN:00,Kahanpaa:ism:03}. In contrast, those of most planetary
nebulae and isolated Herbig AeBe stars are highly variable and are all
shifted towards longer wavelengths (exemplified in
Fig.~\ref{EP:prof_sws} by class B). Important exceptions to such
classification are the post-AGB stars who emit features from all
classes A, B and C. Furthermore, it should be emphasized that while
class A and C show little to no variation in their profiles, large
difference are present within class B.
\begin{figure}[t!]
\includegraphics[clip,width=6.5cm,height=5cm]{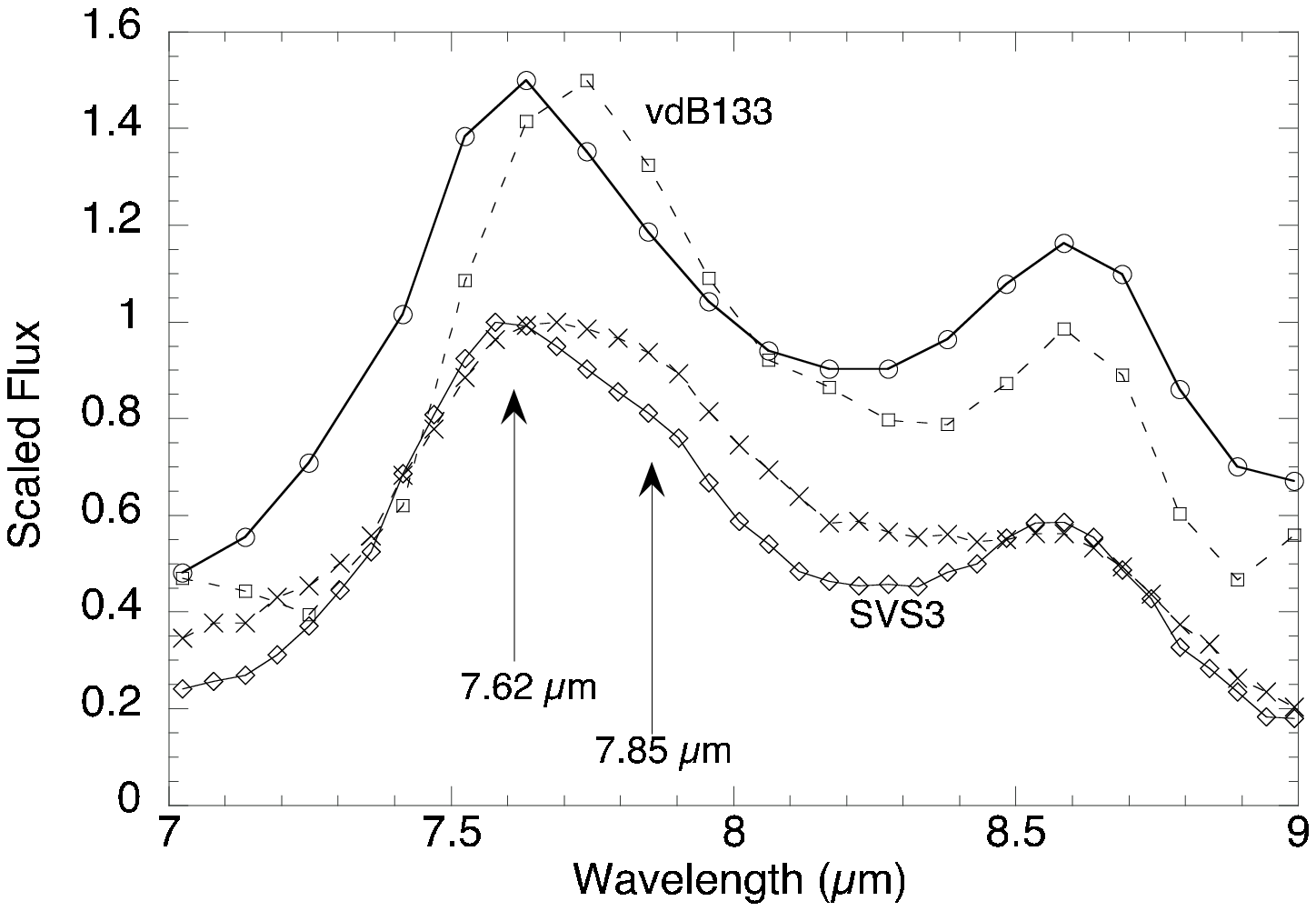}
\hfill
\includegraphics[clip,width=6.5cm,height=5cm]{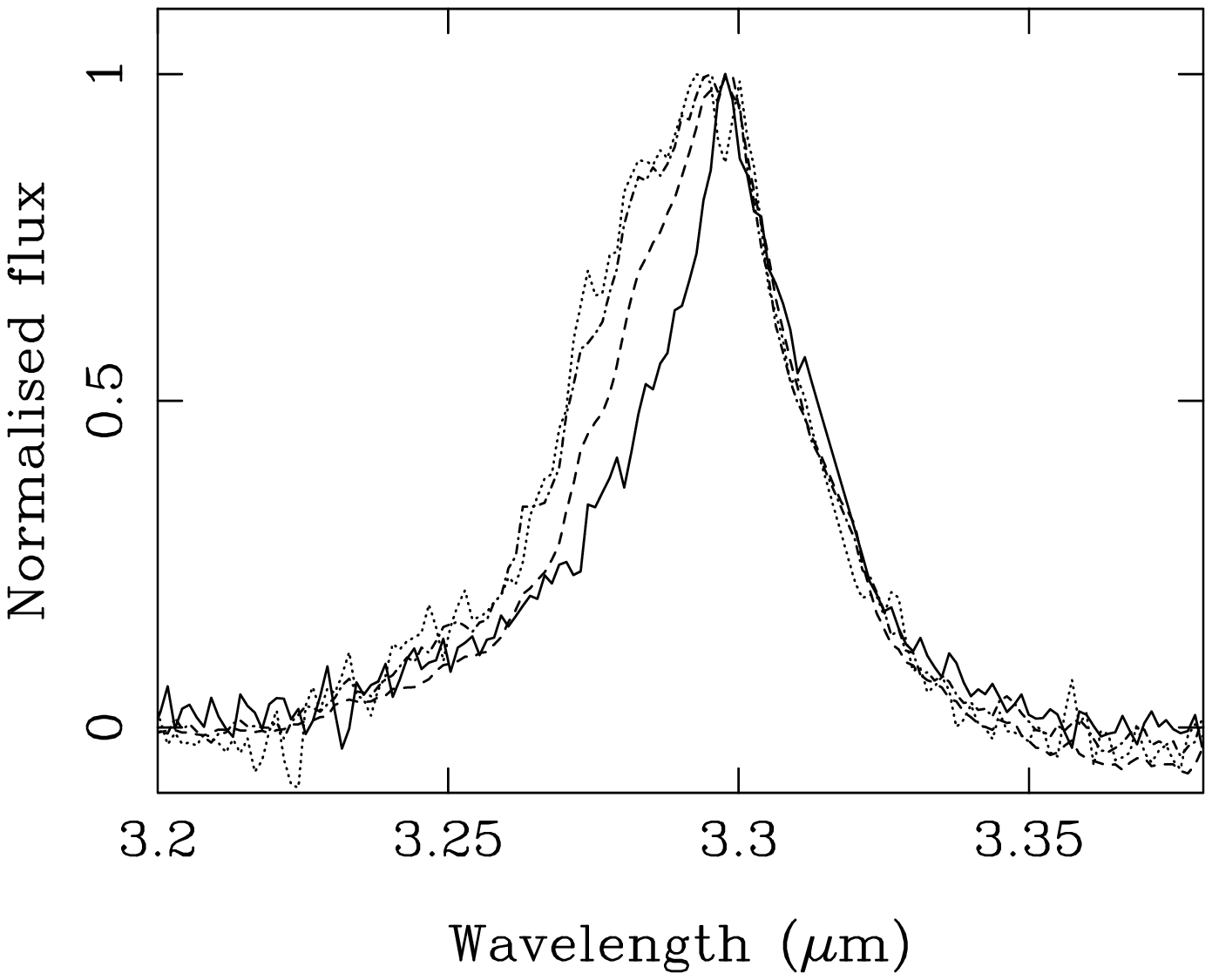}
\caption{ {\bf (left)} The 7.7 and 8.6~$\mu$m UIR features are shown
  for two positions within the reflection nebulae vdB133 (top two
  traces) and SVS3 (lower two traces) obtained with
  ISOCAM-CVF. Courtesy of \citet{Bregman:colorado:03,
  Bregman:04}. {\bf (right)} The observed variations of the 3.3~$\mu$m
  feature in the Red Rectangle with angular distance from the central
  star. The spectra shown are the result of continuum subtraction
  followed by normalization to a common peak intensity. Solid line -
  on-star spectrum; dashed line - 2.6$\arcsec$ spectrum; dot-dash line
  - 3.8$\arcsec$; dotted line - 5$\arcsec$ spectrum. The growth of a
  shoulder with increasing distance from the star is evident. Courtesy
  of \citet{Song:33:03, Song:colorado:03}.}
\label{EP:prof_cam}
\end{figure}

\subsection{Variations within a source}
\label{EP:vari-with-source}
The studies outlined above have concentrated on the variations between
the \emph{entire} integrated UIR spectra from various types of
sources. These studies reveal interesting variations with source type.
It is important to investigate these variations on a smaller spatial
scale as they hold the promise of probing some of the physical
parameters that influence the emergent UIR spectrum.

UIR feature variations within a source were first reported for the
3.3~$\mu$m feature in the Red Rectangle \citep{Kerr:rr:99}.
Approaching the star, the 3.3~$\mu$m profile shifts from a Type~1
profile to a Type~2 profile \citep[Fig.~\ref{EP:prof_cam}, left
  panel][]{Kerr:rr:99, Song:33:03, Song:colorado:03}. Hence close to
the star, the 3.3~$\mu$m feature has the same peak position but a
smaller width than class B while further from the star, they are
similar to those of class A which is associated with
\ion{H}{ii}~regions and the ISM. The 3.3~$\mu$m profile in the ISO-SWS
spectrum (integrated over a 14$\arcsec \times$ 20$\arcsec$ beam)
belongs to class B and hence has a similar peak position as Type 2
features but with a larger FWHM.  The latter is likely caused by the
increasing blue shoulder (as seen in the spatial study), however, it
is unclear why the peak position is not influenced by this.

Similarly, the 7.7 and 8.6~$\mu$m UIR features in the Red Rectangle
show significant spatial variations \citep[][ Song et al. in
preparation]{Song:colorado:03}.  In particular, the 7.7~$\mu$m complex
evolves from a feature with a minor to one with a dominant 7.6~$\mu$m
component with larger distance from the star. Consistent with this,
the 8.6~$\mu$m UIR feature shifts to shorter wavelengths with distance
from the star. Hence, with increasing distance from the star, these
features seem to evolve from those observed in the ISO-SWS
(integrated) spectra of PNe and the Red Rectangle to those of
\ion{H}{ii}~regions and the ISM. In addition, the variations of the
7.7 and 8.6~$\mu$m features within the Red Rectangle show the same
correlated behavior as found from source to source.

\citet{Bregman:colorado:03, Bregman:04} have studied the spatial
behavior of the UIR features in three RNe using ISOCAM-CVF
observations.  ISOCAM allows a spatial resolution of 1.5, 3, 6 or
12$\arcsec$ whereas ISO-SWS spectra integrate a larger area
(14$\arcsec \times$ 20$\arcsec$). Thus, although ISO-SWS spectra of
RNe show similar UIR emission features, these authors found clear
variations in the 7.7~$\mu$m UIR feature (see Fig.~\ref{EP:prof_cam},
right panel). In particular, within vdB133, the profile shifts from
class A$^{\prime}$ to class B$^{\prime}$ in crossing the nebula but in
SVS3 both components are present with variable strength.  The latter
profile is very similar to those in Planetary Nebulae with a a
dominating ``7.8''~$\mu$m peaking at the shortest wavelengths and is
found in regions closer to the illuminating star.

\subsection{Implications}
These observations suggest that in all sources variations in the
profile and peak position occur on a small spatial scale. In addition,
- in contrast to what analysis of the integrated spectra of sources
might suggest - the specific profiles are not unique to certain object
types. In particular, it should be noted that the profiles positioned
at longest wavelengths (class B and Type 2) are found close to the
illuminating star for the three sources (2 RNe and 1 post-AGB star)
discussed above. With increasing distance from the illuminating star,
the UIR feature profiles are more like those found in the integrated
spectra of \ion{H}{ii}~regions and the ISM.  Thus, while the average
(or predominant) profile present within a source does depend on object
type, the variations within sources demonstrate that this likely
reflects the reaction of the PAH family to the very local physical
conditions rather than a difference in chemical history. Thus, the
profile variations between source types reflect the systematic
variations in physical conditions with source type.

The observed variations in the peak position and profile of the
various UIR features provide direct clues to the characteristics of their
carriers and their reaction to a changing environment. Several
different properties/characteristics are proposed to explain them.
Amongst these are anharmonicity, different subcomponents with variable
strength, substituted/complexed PAHs, isotope variations, clustering,
\ldots \, \citep[][Hudgins \& Allamandola, these
proceedings]{Verstraete:m17:96, Pech:prof:01, Verstraete:prof:01,
  Peeters:prof6:02, Wada:13c:03, Song:33:03, Bregman:04,
  vanDiedenhoven:chvscc:03}. Here we will focus on two observational
facts and their implications within the framework of the PAH model --
the (asymmetric) profiles and the 6.2~$\mu$m UIR feature.

\subsubsection{The (asymmetric) profiles}
\label{EP:asymmetric-profiles}
\citet{Barker:anharm:87} has proposed a model to explain the observed
(red shaded) profiles of the 3.3, 6.2, and 11.3 $\mu$m UIR features.
In this interpretation, the red shading comes from the emission from
highly vibrationally excited PAHs. In this case, emission from levels
above the first excited state become important.  Due to the anharmonic
nature of the potential well, these feature spacings become smaller
the higher up the vibrational ladder one samples and emission between
these levels falls progressively to the red producing a red wing
reminiscent of the observed wing. In addition, anharmonic coupling of
the emitting mode with other modes also shifts the peak position of
the emitting feature to lower energies. Integrating over the energy
cascade as the emitting species cools down will then, in a natural
way, give rise to a red shaded profile.  Recently,
\citet{Pech:prof:01} and \citet{Verstraete:prof:01} have modeled the
IR emission spectrum of PAHs based upon extensive laboratory studies
of the shift in peak position as a function of temperature of the
emitter which is a direct measure of the anharmonicity
\citep{Joblin:T:95}.  They obtained excellent fits to the observed
profiles of the 3.3, 6.2 and 11.2~$\mu$m features.

The UIR feature profiles in integrated spectra of many sources are
very similar, irrespective of the illuminating radiation field. As
exemplified by the model study of \citet{Verstraete:prof:01}, this
indicates that the typical size of the emitting PAH is larger in
regions which are illuminated by hotter stars. This coupling between
size and the "color" of the illuminating radiation field may be a
natural consequence of emission from a smooth size distribution of PAH
species; that is, the smallest PAHs which can survive in a radiation
field will depend on the average photon energy in the illuminating FUV
field \citep{Allamandola:rev:89}.

\subsubsection{The variation in the peak position of the 6.2~$\mu$m
  UIR feature}
\label{EP:vari-peak-posit}
Laboratory and theoretical studies show that PAH cations have strong
CC modes longwards of 6.3~$\mu$m with the largest PAH cations emitting
near 6.3~$\mu$m. This behavior is consistent with the limit of
graphite which shows an emission mode at 6.3~$\mu$m
\citep{Draine:graphitegrains:84}. As a consequence, such species
cannot explain a peak position as short as 6.2~$\mu$m. Possibly, this
shift in the peak position of the 6.2~$\mu$m UIR feature results from the
substitution of N into the internal PAH rings in the ISM, although the
chemical processing involved is presently unclear (for a detailed
discussion, see Hudgins \& Allamandola, these proceedings). Another
mechanism proposed to explain these variations is a variation in the
isotopic ratio, $^{12}$C/$^{13}$C \citep{Wada:13c:03}.

\section{Relative intensity variations}
\label{EP:intens}
Correlations between the observed intensities of the major UIR
features based upon KAO data established the notion of a generic UIR
spectrum, which did not vary within the error bars in the (small)
sample investigated \citep[e.g.][]{Cohen:co:86,Cohen:southerniras:89}.
This absence of such variations has sometimes been used as an argument
that the carriers cannot be part of an extended PAH family whose
detailed composition necessarily must vary when the physical
conditions change \citep{Tokunago:97, Boulanger:98}.  Some variations
among the features were however recognized in ground-based data;
notably, the weak 3.4~$\mu$m feature was shown to vary relative to the
main 3.3~$\mu$m feature
\citep{Geballe:89,Moorhouse:89,Joblin:3umvsmethyl:96,Sloan:97} and the
11.2~$\mu$m feature was also shown to vary relative to (the wing of)
the 7.7~$\mu$m feature \citep{Joblin:3umvsmethyl:96,Witteborn:89}.  As
for the peak positions and profiles, the launch of ISO has provided
wide spectral coverage with a single instrument of a large and diverse
set of objects at high signal-to-noise.  This has allowed a detailed
re-examination of the issue of variations in the relative intensities
of the UIR features and this has lead to a reversal of this earlier
conclusion.  It is now clear that the relative strength of the UIR
features varies from source to source and that this provides a tool to
determine the characteristics of the emitting population and,
eventually, to study the physical conditions in the emitting regions.

\subsection{Source to source variations}
\begin{figure}[t!]
\includegraphics[width=6.5cm,height=6.1cm]{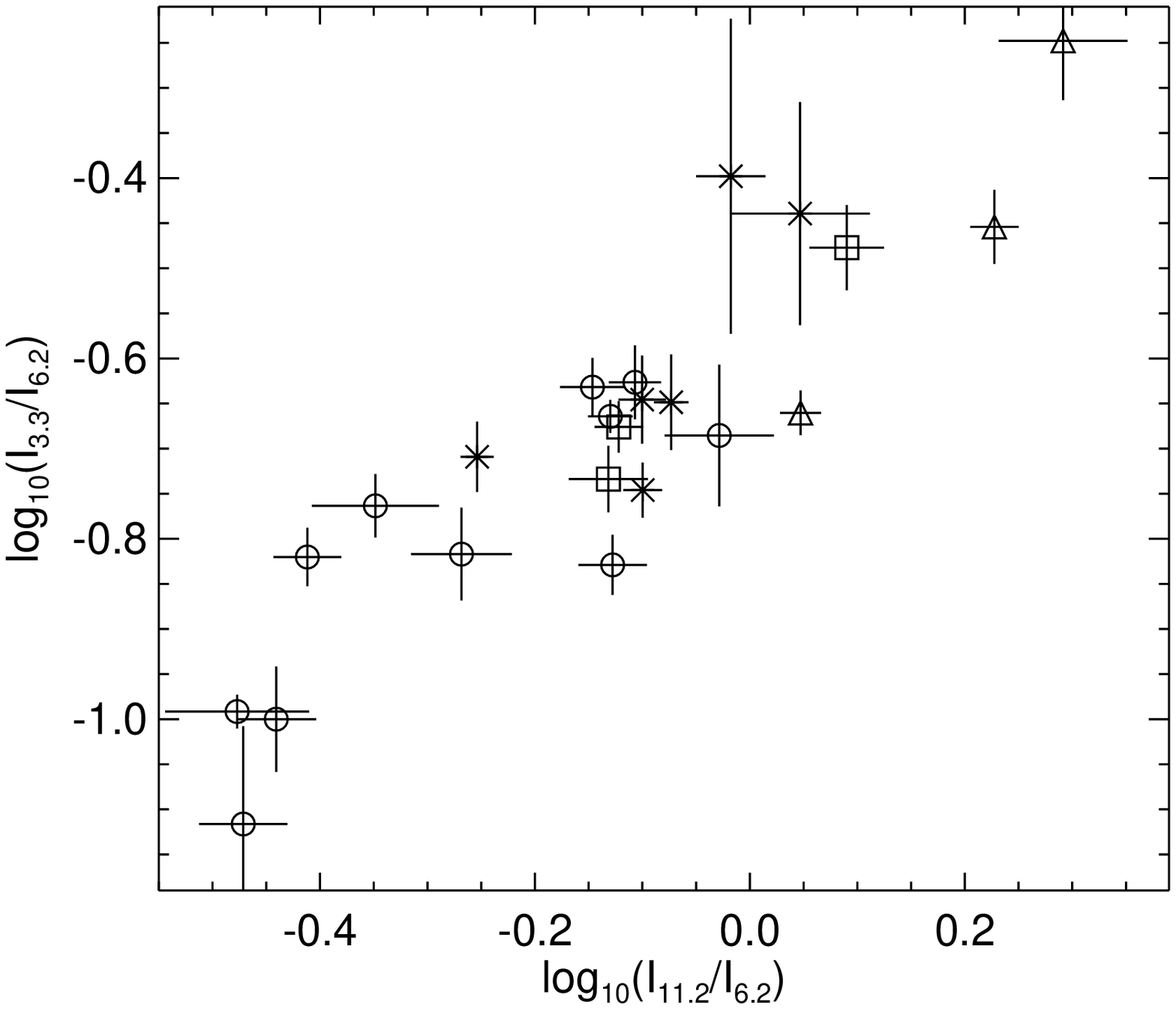}
\hfill
\includegraphics[width=6.5cm,height=6.1cm]{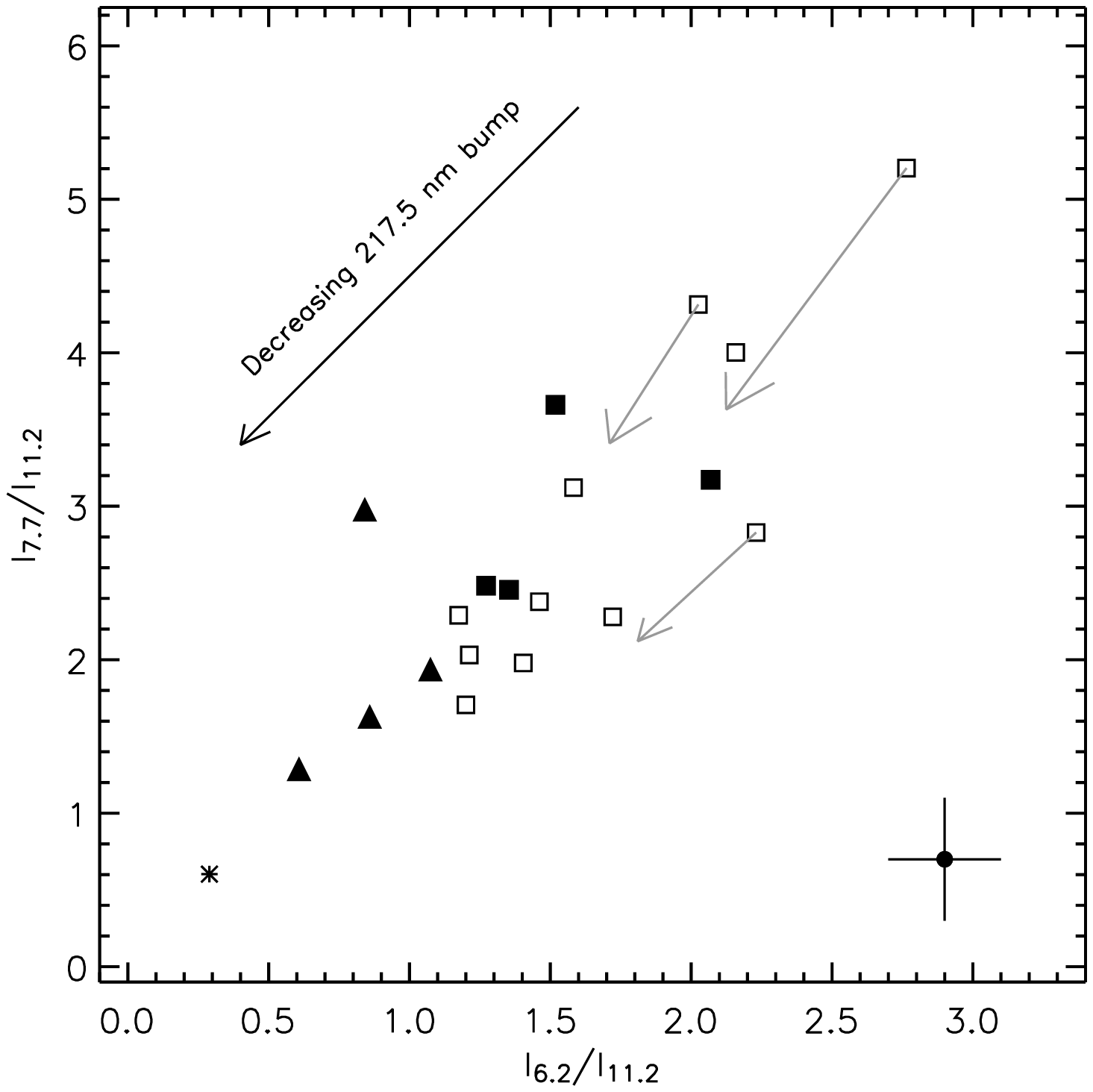}
\caption{{\bf (left)} The I$_{11.2}$/I$_{6.2}$ versus the
I$_{3.3}$/I$_{6.2}$ feature strength ratios as derived from the
ISO-SWS spectra of various sources. Planetary nebulae are represented
by $\triangle$, \ion{H}{ii}~regions by $\circ$, reflection nebulae by
$\Box$ and star forming regions by $\star$. Courtesy of
\citet{Hony:oops:01}. {\bf (right)} The I$_{6.2}$/I$_{11.2}$ versus
I$_{7.7}$/I$_{11.2}$ feature strength ratios as derived from the
ISO-SWS and ISO-PHOT spectra of \ion{H}{ii}~regions within the Galaxy
and the Magellanic Clouds. The Galactic sources are shown as $\Box$,
the non-30~Dor sources as $\blacksquare$ and 30 Dor pointings as
$\blacktriangle$. The asterisk represents the SMC B1$\sharp$1
molecular cloud \citep{Reach:pahsinsmc:00}. Courtesy of
\citet{Vermeij:pahs:01}.}
\label{EP:ratios}
\end{figure}
\begin{figure}[t!]
\includegraphics[clip,width=7.0cm]{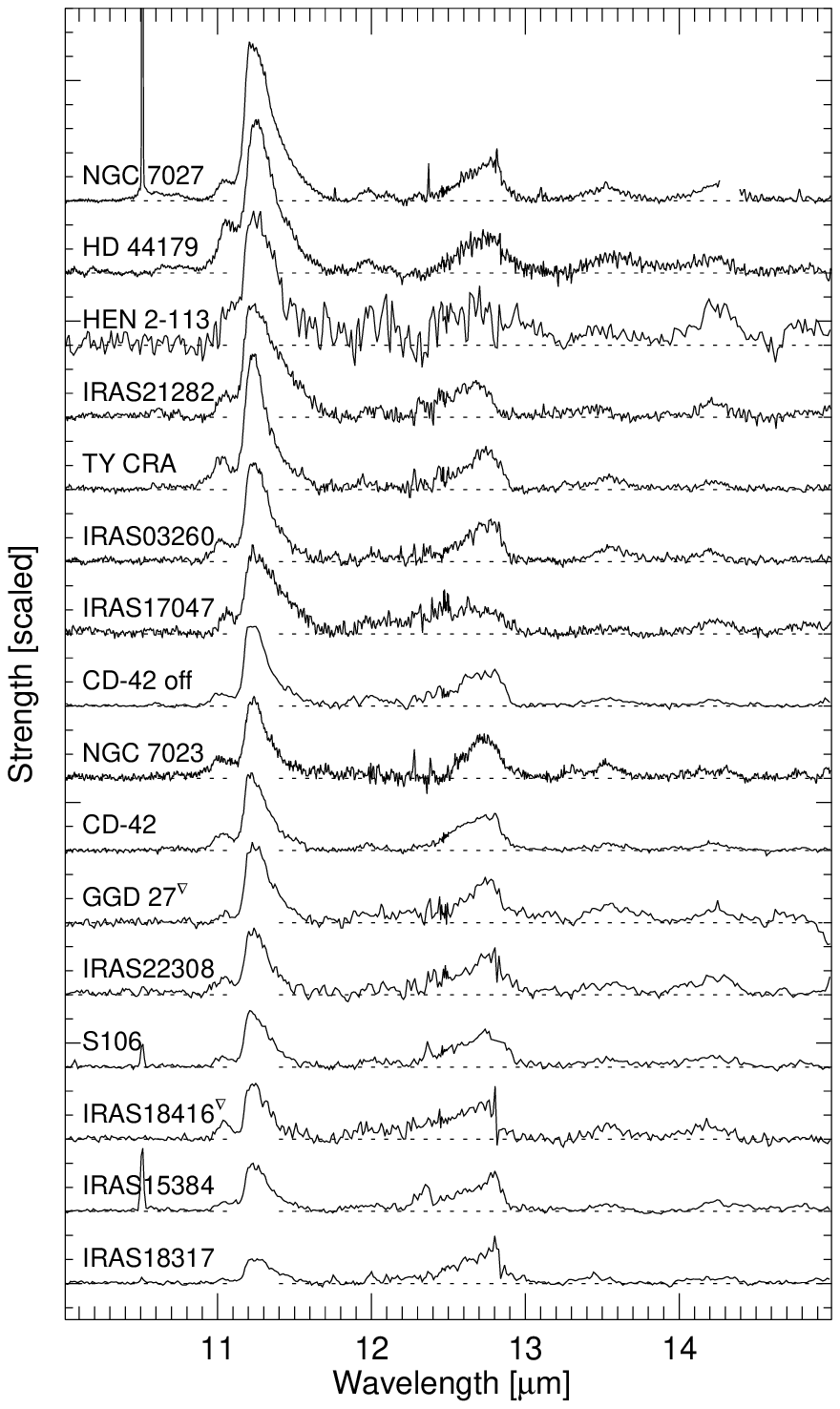}
\hfill
\includegraphics[clip,angle=90,width=5.0cm]{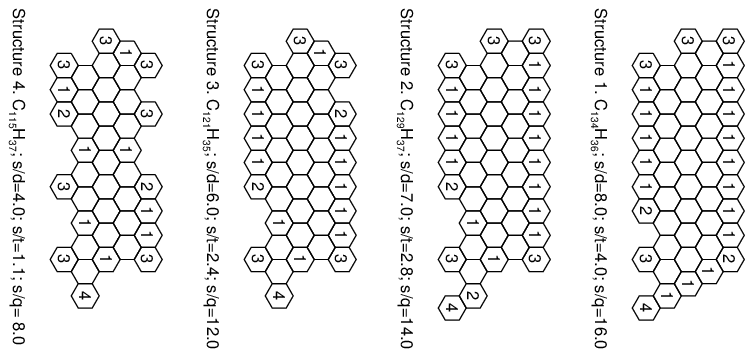}
\caption{{\bf (left)} An overview of the observed features variations in the
out-of-plane bending region. The ISO-SWS spectra are continuum subtracted and
scaled to have the same integrated intensity in the 12.7~$\mu$m UIR
feature. {\bf (right)} Examples of molecular structures simultaneously
satisfying the structural constraints set by the observed feature
strength ratios of the number of solo, duo and trio modes for
different interstellar regions.  The number of solo, duo, trio and
quartet functional groups are noted s, d, t and q respectively. Solo
modes are associated with long straight molecular edges while duos and
trios correspond to corners. Courtesy of \citet{Hony:oops:01}. }
\label{EP:oops}
\end{figure}
In their study of the 11--15~$\mu$m UIR features, \citet{Hony:oops:01}
discovered that -- while all features loosely go together -- there are
variations in the relative strength of the CC modes in the 6--9~$\mu$m
range relative to the CH modes at 3.3 and 11.2~$\mu$m.  Specifically,
the 3.3~$\mu$m feature correlates quite well with the 11.2~$\mu$m
feature but they vary by about a factor 5 relative to the 6.2~$\mu$m
feature (cf., Fig.~\ref{EP:ratios}, left panel).  Likewise, while the
6.2~$\mu$m feature correlates well with the 7.7~$\mu$m feature, both
features vary relative to the CH features.  This can be directly
inferred from the spectra shown in Fig.~\ref{EP:rich}.  These
variations are directly related to object type; that is, while
post-AGB objects and PNe are characterized by high
I$_{11.2}$/I$_{6.2}$ and I$_{3.3}$/I$_{6.2}$ ratios, YSOs, reflection
nebulae, and \ion{H}{ii}~regions are displaced to lower values.  Thus,
the relative intensities of the UIR features reflect the local
physical conditions and/or the local history of the emitting
population.  While a dichotomy between the CH and CC modes exists for
the four major UIR features, this does not extend to all CH
out-of-plane bending modes.  Both the 11.2 and 12.7~$\mu$m features
corresponds to CH out-of-plane bending modes but their relative strength
varies considerably between sources (Fig.~\ref{EP:oops}, left panel).

\subsection{Variation with environment}
\begin{figure}[t!]
\includegraphics[clip,angle=90,width=\textwidth]{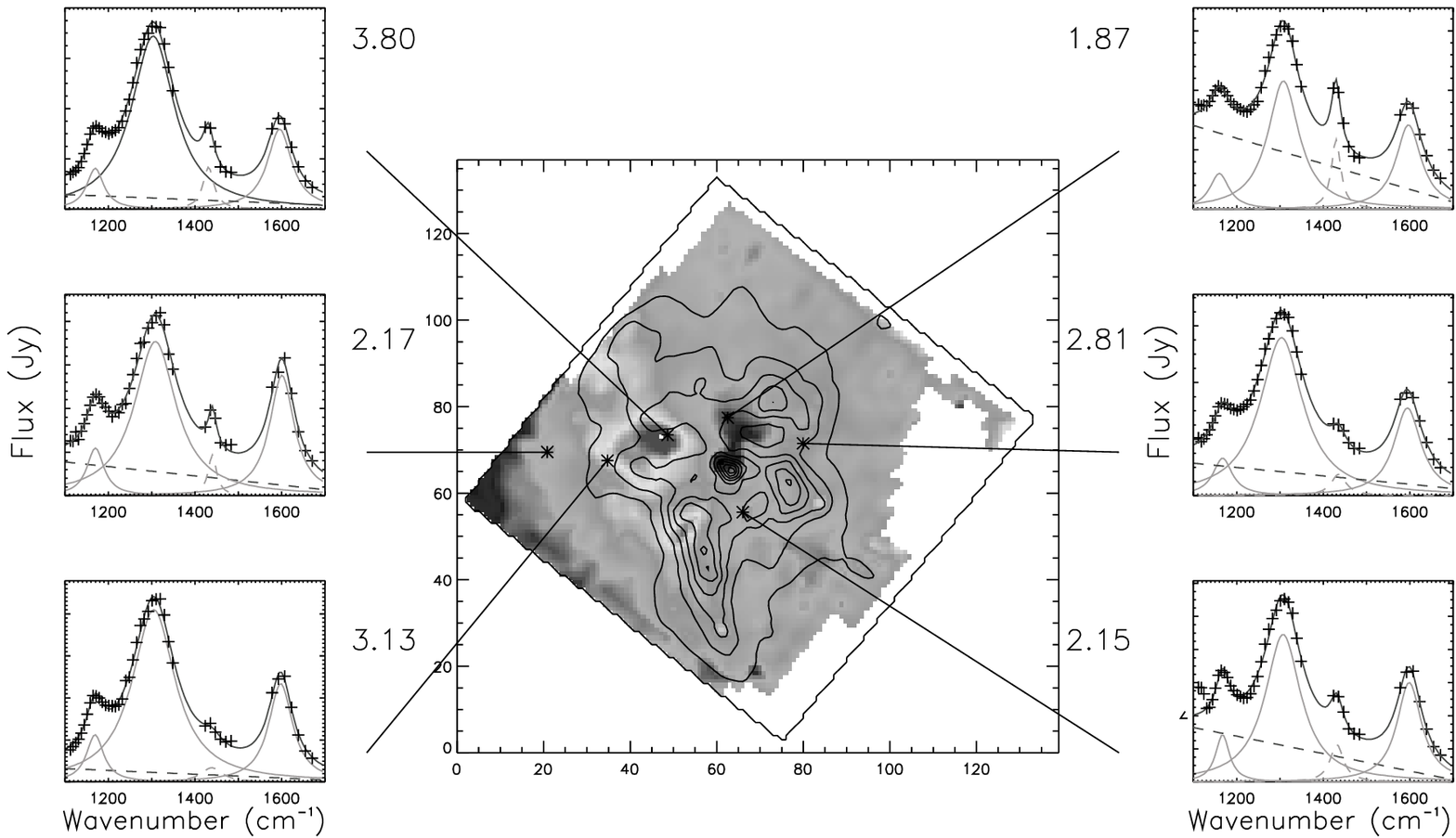}
\caption{Emission in the 7.7~$\mu$m UIR feature (contours) and the
  I$_{7.7}$/I$_{6.2}$ UIR feature intensity ratio (grey scale)
  measured in the \ion{H}{2} region S106 with ISOCAM-CVF. The contours
  clearly show the central exciting source S106-IR and several large
  condensations. The spectrum measured at several positions is also
  displayed. The values on the side give the integrated feature ratio
  of the 7.7 and 6.2~$\mu$m UIR features (at 1299 and 1613 cm$^{-1}$
  respectively). Courtesy of \citet{Joblin:isobeyondthepeaks:00}.}
\label{EP:joblin}
\end{figure}

\citet{Vermeij:pahs:01} studied \ion{H}{ii}~regions in the Milky Way
and the Magellanic Clouds. Consistent with Galactic observations,
these authors find correlations between the I$_{7.7}$/I$_{11.2}$
versus I$_{6.2}$/I$_{11.2}$ and the I$_{8.6}$/I$_{11.2}$ versus
I$_{6.2}$/I$_{11.2}$ feature strength ratios (Fig.~\ref{EP:ratios},
right panel). These plots suggest an interesting segregation between
the sources in the different types of environment (Milky Way -- LMC --
SMC). Furthermore, within the LMC observations, a clear distinction
between 30~Doradus and non-30~Doradus pointings is found. In fact, a
similar fraction of the total PAH energy is emitted in the 7.7 and
8.6~$\mu$m UIR features for the Galactic and Magellanic Cloud
sources. In contrast, the fraction of the total PAH flux emitted in
the 6.2~$\mu$m feature is larger whereas that emitted in
the 11.2~$\mu$m feature is slightly smaller in the Galactic
and non-30~Doradus sources compared to 30~Doradus pointings and the
SMC source. We note that the observed variations in the
I$_{7.7}$/I$_{11.2}$ and I$_{6.2}$/I$_{11.2}$ ratios parallel a
variation in the strength of the 2175 \AA\ bump in the UV extinction
curve \citep{Vermeij:pahs:01}.

\subsection{Variations within sources}
The intensity studies outlined above have concentrated on the
variations between the \emph{entire} integrated UIR spectra from
various types of sources, revealing interesting variations with source
type and environment. As mentioned earlier, the influence of the local
physical conditions (the local radiation field, G$_0$; G$_0$/n$_e$
which determines the charge balance, \ldots) are better addressed
studying variations within a single source or region of the sky.

It is worth noting that the observed intensity variations in
the integrated spectra are necessarily a lower limit to the variations
when studied at higher angular resolution. In the simplest case, this
is due to the fact that the integrated spectrum is weighted to the
brightest UIR emitting regions and any variations in fainter regions
are veiled. In the general case, the different emission features might
arise from different locations within the object but still give rise
to a ``typical'' UIR emission spectrum in the integrated spectrum.

As noted above, the existence of variations among the (minor) features
within a single source were already established earlier from
ground-based observations.  The ISOCAM-CVF instrument has allowed to
extend the search for such variations to the main CC modes in the
5-8~$\mu$m range.  This is illustrated in Fig.~\ref{EP:joblin} for the
bipolar \ion{H}{ii}~region, S106 \citet{Joblin:isobeyondthepeaks:00}.
These spectra reveal variations in the relative strength of
the 6.2 and 7.7~$\mu$m features in this source, which these authors
link to variations in the molecular composition. In particular, high
I$_{7.7}$/I$_{6.2}$ ratios seem to correspond to clumps of CO emission
in the molecular cloud where fresh material is exposed to the strong
radiation field \citep{Joblin:isobeyondthepeaks:00}. These variations
seem to be accompanied by a variation in the strength of the UIR
features relative to the local continuum. Such a variation in the
strength of the CC UIR features relative to the 6~$\mu$m continuum has
also been noted in the \ion{H}{ii}~region M17, on a scale which
allowed ``mapping'' using the
14$^{\prime\prime}$$\times$20$^{\prime\prime}$ aperture of the SWS
instrument \citep{Verstraete:m17:96}.

\subsection{Lack of variations}
\label{EP:lack}
The previously described studies of UIR feature ratios changing within
sources and from source to source, generally refer to regions of high
radiation density.  Most studies of regions of {\it low} radiation
density, however, conclude that there is a striking lack of variations
in the relative intensities of the UIR features in these environments
of low illumination. Most notable are the studies performed with the
\emph{Infrared Telescope in Space} (IRTS) and ISO of the diffuse ISM
\citep[][Onaka, these proceedings]{Onaka:00, Chan:01} and within
reflection nebulae \citep{Uchida:RN:00}. On the average, the relative
strength of the UIR features in these nebulae is insensitive to the
density of incident radiation field (G$_0$). However, these UIR
feature ratios show a large spread around these average, and this
spread may reflect real variation rather than noise. A recent study on
three reflection nebulae by \citet{Bregman:04} reveals variations
in the relative strength of the UIR features with G$_0$/n$_e$, which
determines the charge balance \citep[e.g.][Bakes, these
proceedings]{Bakes:modelI:01}.

\subsection{Implications}
\label{EP:implications}
The observed variations in the relative strength of the various UIR
features provide direct clues to the characteristics of their carriers
and their reaction to a changing environment.  While this analysis is
not yet complete, thanks to a large and dedicated laboratory and
theoretical effort, important information is starting to emerge.  Here
we will focus on two variations and their implications within the
framework of the PAH model -- the variations in the ratio of the CH to
the CC modes feature and the variations in the 11-15~$\mu$m range.

\subsubsection{The ratio of the CH to CC mode intensity and the charge
 balance of PAHs}
\label{EP:11.26.2-mum-ratio}

Laboratory studies and quantum mechanical calculations have shown that
the CC modes are intrinsically weak in neutrals but strong in ions,
while this is reversed for neutrals \citep[][Hudgins \& Allamandola,
these proceedings]{Ellinger:92,Hudgins:tracesionezedpahs:99,
Kim:gasphasepyrenecation:01, Piest:01, Oomens:01,Oomens:03}.  Hence,
the observed variations in the strength of the CH modes relative to
the CC modes may thus reflect a variation in the degree of ionization
of the PAHs in these different environments \citep{Joblin:ngc1333:96,
Hony:oops:01}.  This is supported by detailed model calculations of
the emission spectra of a family of PAHs, taking realistic molecular
properties into account \citep[][and these
proceedings]{Bakes:modelI:01,Bakes:modelII:01}.  In this respect, the
near constant band-strength ratios for a large range of G$_0$ under
low excitation conditions (Sect.~\ref{EP:lack}), seem
puzzling. However, \citet{Bakes:modelI:01} found that the ratios of
the infrared emission features do not vary much for G$_0$/n$_e$ (with
n$_e$ the electron density), which determines the PAH charge and which
is also difficult to determine observationally.  In addition, a
seemingly constant ratio over a large range of G$_0$ would imply a
concerted change in electron density (n$_e$).  Such coupled variations
in G$_0$ and n$_e$ are found in reflection nebulae \citep{YoungOwl:02}.

\subsubsection{The 11-15~$\mu$m CH modes and the molecular structure
  of PAHs}

Extensive laboratory studies have shown that the peak wavelength of
the CH out-of-plane bending mode depends sensitively on the number of
adjacent peripheral C-atoms bonded to a H-atom \citep[][these
proceedings]{Bellamy:58,Hudgins:tracesionezedpahs:99}.  While the
details depend slightly on the charge state of the carrier, the
11.2~$\mu$m feature can be safely ascribed to lone CH groups (e.g. no
H-atoms attached to adjacent C-atoms) and the 12.7~$\mu$m feature
belongs to a trio's (three adjacent C-atoms each with a H attached).
Now, the lone CH groups are part of long straight edges, while trio's
are a characteristic of corners in the molecular structure.
Variations in the relative strength of the 11.2 to the 12.7~$\mu$m
feature are thus indicative of variations in the molecular structure
of the emitting PAHs.  Taking the intrinsic relative strength of these
modes into account, possible PAH structures, corresponding to the
observed interstellar I$_{11.2}$/I$_{12.7}$ feature ratios, are
illustrated in Fig.~\ref{EP:oops} (right panel).  These variations
show that circumstellar PAHs associated with PNe are characterized by
very compact molecular structures.  On the other hand, interstellar
PAHs have more corners either because they are (on average) much
smaller or more irregular larger moieties.

\section{UIR features as a diagnostic tool}
\label{EP:tool}
Since the UIR features originate from such a wide variety of different
types of sources and the feature carriers encounter many different
astronomical environments, they serve as important probes of these
very different emission zones.  For example, the presence and strength
of the UIR features are generally thought to trace star formation on a
Galactic scale.  Since the UIR features in extragalactic
environments are very similar to those in the Galaxy, they also serve
as templates to interpret extragalactic observations.

Concerning galaxies, they are used as qualitative and quantitative
diagnostics of the physical processes powering Galactic nuclei
\citep[e.g.][]{Genzel:ULIRGs:98, Lutz:ULIRGs:98, Mirabel:98,
Charmandaris:99, Rigopoulou:99, Clavel:00, Helou:normalgal:00,
Laurent:00, Tran:01, Peeters:pahtracer:04}.  The UIR line-to-continuum
ratio is on average an order of magnitude smaller for AGNs than for
starburst galaxies.  This, in combination with emission lines and the
dust continuum, is used to distinguish between AGNs, starburst
galaxies and heavily obscured galaxies.  For example, using this
reasoning, it is found that Ultraluminous IR Galaxies (ULIRGs) are
mainly powered by starbursts.  Consequently, the strength of the UIR
features can also probe evolutionary effects in other galaxies on a
universal scale.  Furthermore, by studying the UIR features in
low-metallicity environments \citep[e.g.][Galliano et al., in
preparation]{Vermeij:pahs:01, Madden:colorado:03}, they may be tracing
the early evolutionary state of galaxies as well as probing the
elemental evolution in these galaxies. Finally, a recent study on the
use of the UIR features as a quantitative tracer of star formation
activity found that the UIRs may better be suited as a probe of B
stars which dominate the Galactic stellar energy budget than as a
probe of massive star formation \citep{Peeters:pahtracer:04}. Note as
well that the UIR features do not trace ongoing star formation in highly
obscured galaxies.

Recent studies on the detailed characteristics of the UIR spectra in
galaxies also reveal spectral variations similar to those described
above for Galactic objects. The UIR spectrum of galaxies originates
from star forming regions and Photo-Dissociation regions.  Their
Galactic counterparts exhibit UIR features with similar peak positions
and profiles, suggesting conditions in these external galaxies are
comparable.  Although only three galaxies have sufficient S/N in their
ISO-SWS spectra to be analyzed in detail they show 3.3, 6.2, 7.7 and
8.6~$\mu$m UIR features that are much the same as those associated
with \ion{H}{ii}~regions.  However, the 11.2~$\mu$m profile resembles
that of some evolved stars, not \ion{H}{ii}~regions
\citep{Peeters:prof6:02, vanDiedenhoven:chvscc:03}.  On the other hand, the
(relative) intensities of the UIR features in Galactic and
extragalactic objects show clear variations.  Indeed, intensities
between various \ion{H}{ii}~regions in the Milky Way and in the
Magellanic Clouds vary greatly (see Sect.~\ref{EP:intens}).  In
addition, a study of nearby dwarf galaxies shows a similar range of
relative intensities from galaxy to galaxy and within
individual galaxies \citep[][ Galliano et al., in
preparation]{Madden:colorado:03}.

It is clear that the UIR features carry much information on the
conditions of the emitting region. However, exploitation of this tool
requires a systematic study of the spectral characteristics as a
function of the physical conditions in the emitting regions. Such
studies carry the promise of a rich diagnostic tool that may probe, in
particular, star formation throughout the universe.

\section{Summary \& Prospects}
\label{EP:summary}
The UIR emission features dominate the mid-infrared spectrum of almost
all sources - Galactic and extragalactic. The UIR spectra in these
sources are made up of a rich collection of features. Besides the well-known,
most intense UIR features at 3.3, 6.2, 7.7, 8.6, 11.2 and 12.7~$\mu$m,
many discrete weaker features, subcomponents and very broad structures
are observed. It is now firmly established that the detailed
characteristics (intensity, peak position, profile) of the UIR features
vary from source to source and also spatially within extended sources.
In particular, the 6.2 and 7.7~$\mu$m features are noteworthy in this
respect; while the 6.2~$\mu$m feature peak position varies between
6.25--6.3~$\mu$m in many PNe, it shifts to
6.2~$\mu$m in reflection nebulae and \ion{H}{ii}~regions.  Similar
behavior is found for the 7.7~$\mu$m feature which also consists of two
major components, one at 7.6 and one at ``7.8''~$\mu$m.  In this case their
relative strengths vary between objects.  In \ion{H}{ii}~regions and
reflection nebulae, the 7.6~$\mu$m feature dominates while for most
PNe, the ``7.8''~$\mu$m feature takes over.
Remarkably, in some post-AGB objects however, this feature has shifted
to 8.2~$\mu$m. Thus the 6.2~$\mu$m component is correlated
with the 7.6~$\mu$m feature and the 6.25--6.3~$\mu$m feature with the
``7.8''~$\mu$m component.  While the 3.3 and 11.2~$\mu$m features also show
some variations in peak position and profile, these are much less
pronounced and not as tightly correlated with the variations in the 6.2 and
7.7~$\mu$m features.  In addition, while the 3.3 and 11.2~$\mu$m features
correlate tightly, the strength of these features does vary by a factor 5
relative to the 6.2 and 7.7~$\mu$m features. Nonetheless, the relative
strengths of the features on the 11-15~$\mu$m range do vary. These source
to source variations in intensity, peak position and profile are also
observed spatially within sources and hence are not unique to certain
object types.  Thus, these variations trace local conditions and
reflect the state of the emitting population. Consequently, these
variations are emerging as an integral and important part of the UIR
studies. They imply that the emission is carried by a family of
related chemical compounds whose detailed physical and/or chemical
characteristics vary from source to source and within a given source
in reaction to the local physical conditions.

Due to their omnipresence, the UIR features are now commonly used as
tools for probing objects throughout the universe.  They are used as a
global diagnostic to reveal the different nature of galaxies,
i.e. AGN-dominated, starburst-dominated or heavily obscured.  In
addition, they are a tracer for star formation and the evolutionary
effects in galaxies.  On top of this, the detailed UIR spectra which
have been observed in a handful of galaxies also show the same
significant spectral variations seen in Galactic objects.  Thus, these
UIR features also hold the promise to serve as powerful and detailed
probes of extragalactic objects.

While in the past, progress in this area was limited by astronomical
observations, ISO has relocated the astronomical boundary.  Today, the
richness in spectra and particularly the plethora of spectral
variations from source to source and within sources prompt more
questions than can be answered with the current laboratory and
theoretical data. There have been many comparisons between the spectra
of candidate carriers and the astronomical spectra over the years.
Due to the large effort put in theoretical and laboratory studies, the
agreement with the astronomical spectra are now striking.
Nevertheless, thanks to the quality of the new astronomical
observations, important differences become apparent which shed further
light on the interstellar PAH population. Simultaneously, the wealth
of variations observed in the UIR features puts very strong
constraints on the candidate carriers.

As with ISO, SIRTF will overwhelm the astronomical community with UIR
features. Together with current near- and mid-infrared ground-based
instruments, it will probe variations in the spectra of the UIR
features on small spatial scales allowing us to fully constrain the
variations possible among different objects and to determine, in
detail, the interplay of physical conditions, environment and UIR
feature characteristics.  Due to SIRTF's superb sensitivity, the UIR
feature characteristics will also be revealed for a large number of
galaxies.

The observational, theoretical and experimental tools are now in place
and are being further developed to tackle this vast treasure trove of
information which will truly probe most of the universe.

\acknowledgments We gratefully acknowledge Jesse Bregman, Christine
Joblin, Peter Sarre, In-Ok Song who shared their data and preprints
with us.  We also thank MNRAS for permission to reproduce Fig. 3,
right panel. This review wouldn't be the same without ISO, an ESA
project with instruments funded by ESA Member States (especially the
PI countries: France, Germany, the Netherlands and the United Kingdom)
and with the participation of ISAS and NASA. EP acknowledges the
financial support of the National Research Council.


\begin{thebibliography}{110}
\expandafter\ifx\csname natexlab\endcsname\relax\def\natexlab#1{#1}\fi

\bibitem[{{Aannestad} \& {Kenyon}(1979)}]{Aannestad:79}
{Aannestad}, P.~A. \& {Kenyon}, S.~J. 1979, \apj, 230, 771

\bibitem[{{Allain} {et~al.}(1996){Allain}, {Leach}, \&
  {Sedlmayr}}]{Allain:ionendehydro:96}
{Allain}, T., {Leach}, S., \& {Sedlmayr}, E. 1996, \aap, 305, 616

\bibitem[{{Allamandola} {et~al.}(1999){Allamandola}, {Hudgins}, \&
  {Sandford}}]{Allamandola:modelobs:99}
{Allamandola}, L.~J., {Hudgins}, D.~M., \& {Sandford}, S.~A. 1999, \apjl, 511,
  L115

\bibitem[{{Allamandola} {et~al.}(1989){Allamandola}, {Tielens}, \&
  {Barker}}]{Allamandola:rev:89}
{Allamandola}, L.~J., {Tielens}, A.~G.~G.~M., \& {Barker}, J.~R. 1989, \apjs,
  71, 733

\bibitem[{{Bakes} \& {Tielens}(1994)}]{Bakes:photoelec:94}
{Bakes}, E.~L.~O. \& {Tielens}, A.~G.~G.~M. 1994, \apj, 427, 822

\bibitem[{{Bakes} \& {Tielens}(1998)}]{Bakes:98}
{Bakes}, E.~L.~O. \& {Tielens}, A.~G.~G.~M. 1998, \apj, 499, 258

\bibitem[{{Bakes} {et~al.}(2001{\natexlab{a}}){Bakes}, {Tielens}, \&
  {Bauschlicher}}]{Bakes:modelI:01}
{Bakes}, E.~L.~O., {Tielens}, A.~G.~G.~M., \& {Bauschlicher}, C.~W.
  2001{\natexlab{a}}, \apj, 556, 501

\bibitem[{{Bakes} {et~al.}(2001{\natexlab{b}}){Bakes}, {Tielens},
  {Bauschlicher}, {Hudgins}, \& {Allamandola}}]{Bakes:modelII:01}
{Bakes}, E.~L.~O., {Tielens}, A.~G.~G.~M., {Bauschlicher}, C.~W., {Hudgins},
  D.~M., \& {Allamandola}, L.~J. 2001{\natexlab{b}}, \apj, 560, 261

\bibitem[{{Barker} {et~al.}(1987){Barker}, {Allamandola}, \&
  {Tielens}}]{Barker:anharm:87}
{Barker}, J.~R., {Allamandola}, L.~J., \& {Tielens}, A.~G.~G.~M. 1987, \apjl,
  315, L61

\bibitem[{{Bauschlicher} {et~al.}(2004){Bauschlicher}, {Hudgins}, \&
  {Allamandola}}]{Bauschlicher:NPAHs:04}
{Bauschlicher}, J. C.~W., {Hudgins}, D.M., \& {Allamandola}, L.J. 2004, \apj, submitted

\bibitem[{{Bellamy}(1958)}]{Bellamy:58}
{Bellamy}, L. 1958, The infra-red spectra of complex molecules, 2nd ed. (New
  York: John Wiley {\&} Sons, Inc.)

\bibitem[{{Borghesi} {et~al.}(1987){Borghesi}, {Bussoletti}, \&
  {Colangeli}}]{Borghesi:amcarbon:87}
{Borghesi}, A., {Bussoletti}, E., \& {Colangeli}, L. 1987, \apj, 314, 422

\bibitem[{{Boulanger}(1999)}]{Boulanger:leshouches:99}
{Boulanger}, F. 1999, in Solid Interstellar Matter : The ISO Revolution, (eds.)
  d'Hendecourt L., Joblin C., Jones A., EDP Sciences, 20

\bibitem[{{Boulanger} {et~al.}(1998{\natexlab{a}}){Boulanger}, {Abergel},
  {Bernard}, {Cesarsky}, {Puget}, {Reach}, {Ryter}, {Cesarsky}, {Sauvage},
  {Tran}, {Vigroux}, {Falgarone}, {Lequeux}, {Perault}, \&
  {Rouan}}]{Boulanger:98}
{Boulanger}, F., {Abergel}, A., {Bernard}, J.~P., {et~al.} 1998{\natexlab{a}},
  in ASP Conf. Ser. 132: Star Formation with the Infrared Space Observatory, (eds.) 
  Yun J. \&  Liseau R., 15

\bibitem[{{Boulanger} {et~al.}(2000){Boulanger}, {Abergel}, {Cesarsky},
  {Bernard}, {Miville Desch{\^ e}nes}, {Verstraete}, \& {Reach}}]{Boulanger:00}
{Boulanger}, F., {Abergel}, A., {Cesarsky}, D., {et~al.} 2000, ISO Beyond Point
  Sources: Studies of Extended Infrared Emission, (eds.) Laureijs R.~J., Leech K. 
  \& Kessler M.~F., 455, 91

\bibitem[{{Boulanger} {et~al.}(1998{\natexlab{b}}){Boulanger}, {Boisssel},
  {Cesarsky}, \& {Ryter}}]{Boulanger:lorentz:98}
{Boulanger}, F., {Boisssel}, P., {Cesarsky}, D., \& {Ryter}, C.
  1998{\natexlab{b}}, \aap, 339, 194

\bibitem[{{Bregman}(1989)}]{Bregman:hiivspn:89}
{Bregman}, J. 1989, in IAU Symp. 135: Interstellar Dust, (eds.) Allamandola L.~J. 
  \& Tielens A.~G.~G.~M., 109

\bibitem[{{Bregman} \& {Temi}(2003)}]{Bregman:colorado:03}
{Bregman}, J. \& {Temi}, P. 2003, in Astrophysics of Dust, (ed.) Witt A.~N.

\bibitem[{{Bregman} \& {Temi}(2004)}]{Bregman:04}
{Bregman}, J. \& {Temi}, P. 2004, \aap, submitted

\bibitem[{{Bregman} {et~al.}(1989){Bregman}, {Allamandola}, {Witteborn},
  {Tielens}, \& {Geballe}}]{Bregman:orion:89}
{Bregman}, J.~D., {Allamandola}, L.~J., {Witteborn}, F.~C., {Tielens},
  A.~G.~G.~M., \& {Geballe}, T.~R. 1989, \apj, 344, 791

\bibitem[{{Chan} {et~al.}(2001){Chan}, {Roellig}, {Onaka}, {Mizutani},
  {Okumura}, {Yamamura}, {Tanab{\' e}}, {Shibai}, {Nakagawa}, \&
  {Okuda}}]{Chan:01}
{Chan}, K., {Roellig}, T.~L., {Onaka}, T., {et~al.} 2001, \apj, 546, 273

\bibitem[{{Charmandaris} {et~al.}(1999){Charmandaris}, {Laurent}, {Mirabel},
  {Gallais}, {Sauvage}, {Vigroux}, {Cesarsky}, \& {Tran}}]{Charmandaris:99}
{Charmandaris}, V., {Laurent}, O., {Mirabel}, I.~F., {et~al.} 1999, \apss, 266,
  99

\bibitem[{{Cherchneff} {et~al.}(2000){Cherchneff}, {Le Teuff}, {Williams}, \&
  {Tielens}}]{Cherchneff:dustformwr:00}
{Cherchneff}, I., {Le Teuff}, Y.~H., {Williams}, P.~M., \& {Tielens},
  A.~G.~G.~M. 2000, \aap, 357, 572

\bibitem[{{Chiar} {et~al.}(2002){Chiar}, {Peeters}, \& {Tielens}}]{Chiar:wr:02}
{Chiar}, J.~E., {Peeters}, E., \& {Tielens}, A.~G.~G.~M. 2002, \apjl, 579, L91

\bibitem[{{Clavel} {et~al.}(2000){Clavel}, {Schulz}, {Altieri}, {Barr},
  {Claes}, {Heras}, {Leech}, {Metcalfe}, \& {Salama}}]{Clavel:00}
{Clavel}, J., {Schulz}, B., {Altieri}, B., {et~al.} 2000, \aap, 357, 839

\bibitem[{{Cohen} {et~al.}(1986){Cohen}, {Allamandola}, {Tielens}, {Bregman},
  {Simpson}, {Witteborn}, {Wooden}, \& {Rank}}]{Cohen:co:86}
{Cohen}, M., {Allamandola}, L., {Tielens}, A.~G.~G.~M., {et~al.} 1986, \apj,
  302, 737

\bibitem[{{Cohen} {et~al.}(1989){Cohen}, {Tielens}, {Bregman}, {Witteborn},
  {Rank}, {Allamandola}, {Wooden}, \& {Jourdain de
  Muizon}}]{Cohen:southerniras:89}
{Cohen}, M., {Tielens}, A.~G.~G.~M., {Bregman}, J., {et~al.} 1989, \apj, 341,
  246

\bibitem[{Cox \& Kessler(1999)}]{isoparijs}
Cox, P. \& Kessler, M., eds. 1999, The Universe as Seen by ISO, ESA-SP 427

\bibitem[{{Cr{\' e}t{\' e}} {et~al.}(1999){Cr{\' e}t{\' e}}, {Giard}, {Joblin},
  {Vauglin}, {L{\' e}ger}, \& {Rouan}}]{Crete:99}
{Cr{\' e}t{\' e}}, E., {Giard}, M., {Joblin}, C., {et~al.} 1999, \aap, 352, 277

\bibitem[{{Draine}(1984)}]{Draine:graphitegrains:84}
{Draine}, B.~T. 1984, \apjl, 277, L71

\bibitem[{{Draine}(2003)}]{Draine:dust:03}
{Draine}, B.~T. 2003, in Carnegie Observatories Astrophysics Series, Vol. 4 :
  Origin and Evolution of the Elements, (eds) A. McWilliam and M. Rauch, in press

\bibitem[{{Duley} \& {Williams}(1981)}]{Duley:aromatic:81}
{Duley}, W.~W. \& {Williams}, D.~A. 1981, \mnras, 196, 269

\bibitem[{{Duley} \& {Williams}(1983)}]{Duley:83}
{Duley}, W.~W. \& {Williams}, D.~A. 1983, \mnras, 205, 67P

\bibitem[{{Dwek}(1986)}]{Dwek:86}
{Dwek}, E. 1986, \apj, 302, 363

\bibitem[{{Ellinger}(1992)}]{Ellinger:92}
{Ellinger}, Y. 1992, in IAU Symp. 150: Astrochemistry of Cosmic Phenomena, (ed.) Singh P.~D., 31

\bibitem[{{Frenklach} \& {Feigelson}(1989)}]{Frenklach:form:89}
{Frenklach}, M. \& {Feigelson}, E.~D. 1989, \apj, 341, 372

\bibitem[{{Geballe} {et~al.}(1985){Geballe}, {Lacy}, {Persson}, {McGregor}, \&
  {Soifer}}]{Geballe:85}
{Geballe}, T.~R., {Lacy}, J.~H., {Persson}, S.~E., {McGregor}, P.~J., \&
  {Soifer}, B.~T. 1985, \apj, 292, 500

\bibitem[{{Geballe} {et~al.}(1989){Geballe}, {Tielens}, {Allamandola},
  {Moorhouse}, \& {Brand}}]{Geballe:89}
{Geballe}, T.~R., {Tielens}, A.~G.~G.~M., {Allamandola}, L.~J., {Moorhouse},
  A., \& {Brand}, P.~W.~J.~L. 1989, \apj, 341, 278

\bibitem[{{Genzel} {et~al.}(1998){Genzel}, {Lutz}, {Sturm}, {Egami}, {Kunze},
  {Moorwood}, {Rigopoulou}, {Spoon}, {Sternberg}, {Tacconi-Garman}, {Tacconi},
  \& {Thatte}}]{Genzel:ULIRGs:98}
{Genzel}, R., {Lutz}, D., {Sturm}, E., {et~al.} 1998, \apj, 498, 579

\bibitem[{{Gillett} {et~al.}(1973){Gillett}, {Forrest}, \&
  {Merrill}}]{Gillett:73}
{Gillett}, F.~C., {Forrest}, W.~J., \& {Merrill}, K.~M. 1973, \apj, 183, 87

\bibitem[{{Harrington} {et~al.}(1998){Harrington}, {Lame}, {Borkowski},
  {Bregman}, \& {Tsvetanov}}]{Harrington:64:98}
{Harrington}, J.~P., {Lame}, N.~J., {Borkowski}, K.~J., {Bregman}, J.~D., \&
  {Tsvetanov}, Z.~I. 1998, \apjl, 501, L123

\bibitem[{{Helou} {et~al.}(2000){Helou}, {Lu}, {Werner}, {Malhotra}, \&
  {Silbermann}}]{Helou:normalgal:00}
{Helou}, G., {Lu}, N.~Y., {Werner}, M.~W., {Malhotra}, S., \& {Silbermann}, N.
  2000, \apjl, 532, L21

\bibitem[{{Holmlid}(2000)}]{Holmlid:rydbergmatter:00}
{Holmlid}, L. 2000, \aap, 358, 276

\bibitem[{{Hony} {et~al.}(2001){Hony}, {Van Kerckhoven}, {Peeters}, {Tielens},
  {Hudgins}, \& {Allamandola}}]{Hony:oops:01}
{Hony}, S., {Van Kerckhoven}, C., {Peeters}, E., {et~al.} 2001, \aap, 370, 1030

\bibitem[{{Hudgins} \& {Allamandola}(1999)}]{Hudgins:tracesionezedpahs:99}
{Hudgins}, D.~M. \& {Allamandola}, L.~J. 1999, \apjl, 516, L41

\bibitem[{{Joblin} {et~al.}(2000){Joblin}, {Abergel}, {Bregman},
  {D'Hendecourt}, {Heras}, {Jourdain de Muizon}, {Pech}, \&
  {Tielens}}]{Joblin:isobeyondthepeaks:00}
{Joblin}, C., {Abergel}, A., {Bregman}, J., {et~al.} 2000, ISO beyond the
  peaks: The 2nd ISO workshop on analytical spectroscopy. (eds.) Salama A.,
  Kessler M.~F., Leech K. \& Schulz B.~ESA-SP456, 49

\bibitem[{{Joblin} {et~al.}(1995){Joblin}, {Boissel}, {Leger}, {D'Hendecourt},
  \& {Defourneau}}]{Joblin:T:95}
{Joblin}, C., {Boissel}, P., {Leger}, A., {D'Hendecourt}, L., \& {Defourneau},
  D. 1995, \aap, 299, 835

\bibitem[{{Joblin} {et~al.}(1996{\natexlab{a}}){Joblin}, {Tielens}, 
  {Allamandola}, \& {Geballe}}]{Joblin:3umvsmethyl:96}
{Joblin}, C., {Tielens}, A.~G.~G.~M., {Allamandola}, L.~J., \& {Geballe}, T.~R.
  1996{\natexlab{a}}, \apj, 458, 610

\bibitem[{{Joblin} {et~al.}(1996{\natexlab{b}}){Joblin}, {Tielens}, 
  {Geballe}, \& {Wooden}}]{Joblin:ngc1333:96}
{Joblin}, C., {Tielens}, A.~G.~G.~M., {Geballe}, T.~R., \& {Wooden}, D.~H.
  1996{\natexlab{b}}, \apj, 460, L119


\bibitem[{{Jones} \& {d'Hendecourt}(2000)}]{Jones:nanodiamonds:00}
{Jones}, A.~P. \& {d'Hendecourt}, L. 2000, \aap, 355, 1191

\bibitem[{{Kahanp{\" a}{\" a}} {et~al.}(2003){Kahanp{\" a}{\" a}}, {Mattila},
  {Lehtinen}, {Leinert}, \& {Lemke}}]{Kahanpaa:ism:03}
{Kahanp{\" a}{\" a}}, J., {Mattila}, K., {Lehtinen}, K., {Leinert}, C., \&
  {Lemke}, D. 2003, \aap, 405, 999

\bibitem[{{Kerr} {et~al.}(1999){Kerr}, {Hurst}, {Miles}, \&
  {Sarre}}]{Kerr:rr:99}
{Kerr}, T.~H., {Hurst}, M.~E., {Miles}, J.~R., \& {Sarre}, P.~J. 1999, \mnras,
  303, 446

\bibitem[{{Kim} {et~al.}(2001){Kim}, {Wagner}, \&
  {Saykally}}]{Kim:gasphasepyrenecation:01}
{Kim}, H.~S., {Wagner}, D.~R., \& {Saykally}, R.~J. 2001, \prl, 86, 5691

\bibitem[{{Laurent} {et~al.}(2000){Laurent}, {Mirabel}, {Charmandaris},
  {Gallais}, {Madden}, {Sauvage}, {Vigroux}, \& {Cesarsky}}]{Laurent:00}
{Laurent}, O., {Mirabel}, I.~F., {Charmandaris}, V., {et~al.} 2000, \aap, 359,
  887

\bibitem[{{Le Page} {et~al.}(2003){Le Page}, {Snow}, \& {Bierbaum}}]{LePage:03}
{Le Page}, V., {Snow}, T.~P., \& {Bierbaum}, V.~M. 2003, \apj, 584, 316

\bibitem[{{Lepp} \& {Dalgarno}(1988)}]{Lepp:88}
{Lepp}, S. \& {Dalgarno}, A. 1988, \apj, 324, 553

\bibitem[{{Lutz} {et~al.}(1998){Lutz}, {Spoon}, {Rigopoulou}, {Moorwood}, \&
  {Genzel}}]{Lutz:ULIRGs:98}
{Lutz}, D., {Spoon}, H.~W.~W., {Rigopoulou}, D., {Moorwood}, A.~F.~M., \&
  {Genzel}, R. 1998, \apjl, 505, L103

\bibitem[{{Madden} {et~al.}(2003){Madden}, {Galliano}, \&
  {Jones}}]{Madden:colorado:03}
{Madden}, S., {Galliano}, F., \& {Jones}, A. 2003, in Astrophysics of Dust,
  (ed.) Witt A.~N.

\bibitem[{{Mattila} {et~al.}(1996){Mattila}, {Lemke}, {Haikala}, {Laureijs},
  {Leger}, {Lehtinen}, {Leinert}, \& {Mezger}}]{Mattila:96}
{Mattila}, K., {Lemke}, D., {Haikala}, L.~K., {et~al.} 1996, \aap, 315, L353

\bibitem[{{Mirabel} {et~al.}(1998){Mirabel}, {Vigroux}, {Charmandaris},
  {Sauvage}, {Gallais}, {Tran}, {Cesarsky}, {Madden}, \& {Duc}}]{Mirabel:98}
{Mirabel}, I.~F., {Vigroux}, L., {Charmandaris}, V., {et~al.} 1998, \aap, 333,
  L1

\bibitem[{{Molster} {et~al.}(1996){Molster}, {van den Ancker}, {Tielens},
  {Waters}, {Beintema}, {Waelkens}, {de Jong}, {de Graauw}, {Justtanont},
  {Yamamura}, {Vandenbussche}, \& {Heras}}]{Molster:pah:96}
{Molster}, F.~J., {van den Ancker}, M.~E., {Tielens}, A.~G.~G.~M., {et~al.}
  1996, \aap, 315, L373

\bibitem[{{Moorhouse} {et~al.}(1989){Moorhouse}, {Geballe}, \&
  {Allamandola}}]{Moorhouse:89} {Moorhouse}, A., {Geballe}, T.~R., \&
  {Allamandola}, L.~J. 1989, in Interstellar Dust, 107

\bibitem[{{Moutou} {et~al.}(1999{\natexlab{a}}){Moutou}, {Sellgren},
  {L\'{e}ger}, {Verstraete}, \& {Le Coupanec}}]{Moutou:leshouches:99}
{Moutou}, C., {Sellgren}, K., {L\'{e}ger}, A., {Verstraete}, L., \& {Le
  Coupanec}, P. 1999{\natexlab{a}}, in Solid Interstellar Matter : The ISO
  Revolution, (eds.) d'Hendecourt L., Joblin C., Jones A., EDP Sciences, 90

\bibitem[{{Moutou} {et~al.}(1999{\natexlab{b}}){Moutou}, {Sellgren},
  {Verstraete}, \& {L{\' e}ger}}]{Moutou:c60:99}
{Moutou}, C., {Sellgren}, K., {Verstraete}, L., \& {L{\' e}ger}, A.
  1999{\natexlab{b}}, \aap, 347, 949

\bibitem[{{Moutou} {et~al.}(2000){Moutou}, {Verstraete}, {L{\' e}ger},
  {Sellgren}, \& {Schmidt}}]{Moutou:16.4:00}
{Moutou}, C., {Verstraete}, L., {L{\' e}ger}, A., {Sellgren}, K., \& {Schmidt},
  W. 2000, \aap, 354, L17

\bibitem[{{Moutou} {et~al.}(1999{\natexlab{c}}){Moutou}, {Verstraete},
  {Sellgren}, \& {L\'{e}ger}}]{Moutou:parijs:99}
{Moutou}, C., {Verstraete}, L., {Sellgren}, K., \& {L\'{e}ger}, A.
  1999{\natexlab{c}}, in The Universe as Seen by ISO, (eds.) Cox P. \& Kessler M.~F.,
   ESA SP-427, 727

\bibitem[{{Onaka}(2000)}]{Onaka:00}
{Onaka}, T. 2000, Advances in Space Research, 25, 2167

\bibitem[{{Onaka} {et~al.}(1996){Onaka}, {Yamamura}, {Tanabe}, {Roellig}, \&
  {Yuen}}]{Onaka:dism:96}
{Onaka}, T., {Yamamura}, I., {Tanabe}, T., {Roellig}, T.~L., \& {Yuen}, L.
  1996, \pasj, 48, L59

\bibitem[{{Oomens} {et~al.}(2001){Oomens}, {Sartakov}, {Tielens}, {Meijer}, \&
  {von Helden}}]{Oomens:01}
{Oomens}, J., {Sartakov}, B.~G., {Tielens}, A.~G.~G.~M., {Meijer}, G., \& {von
  Helden}, G. 2001, \apjl, 560, L99

\bibitem[{{Oomens} {et~al.}(2003){Oomens}, {Tielens}, {Sartakov}, {von Helden},
  \& {Meijer}}]{Oomens:03}
{Oomens}, J., {Tielens}, A.~G.~G.~M., {Sartakov}, B.~G., {von Helden}, G., \&
  {Meijer}, G. 2003, \apj, 591, 968

\bibitem[{{Papoular} {et~al.}(1989){Papoular}, {Conrad}, {Giuliano}, {Kister},
  \& {Mille}}]{Papoular:coalmodel:89}
{Papoular}, R., {Conrad}, J., {Giuliano}, M., {Kister}, J., \& {Mille}, G.
  1989, \aap, 217, 204

\bibitem[{{Pech} {et~al.}(2002){Pech}, {Joblin}, \& {Boissel}}]{Pech:prof:01}
{Pech}, C., {Joblin}, C., \& {Boissel}, P. 2002, \aap, 388, 639

\bibitem[{{Peeters} {et~al.}(2004{\natexlab{a}}){Peeters}, {Allamandola},
  {Bauschlicher}, {Hudgins}, {Sandford}, \& {Tielens}}]{Peeters:pads:04}
{Peeters}, E., {Allamandola}, L.~J., {Bauschlicher}, C.~W., {et~al.}
  2004{\natexlab{a}}, \apj, submitted

\bibitem[{{Peeters} {et~al.}(2002){Peeters}, {Hony}, {Van Kerckhoven},
  {Tielens}, {Allamandola}, {Hudgins}, \& {Bauschlicher}}]{Peeters:prof6:02}
{Peeters}, E., {Hony}, S., {Van Kerckhoven}, C., {et~al.} 2002, \aap, 390, 1089

\bibitem[{{Peeters} {et~al.}(2004{\natexlab{b}}){Peeters}, {Spoon}, \&
  {Tielens}}]{Peeters:pahtracer:04}
{Peeters}, E., {Spoon}, H.~W.~W., \& {Tielens}, A.~G.~G.~M. 2004{\natexlab{b}},
  \apj, submitted

\bibitem[{{Peeters} {et~al.}(2004{\natexlab{c}}){Peeters}, {Tielens},
  {Boogert}, {Hayward}, \& {Allamandola}}]{Peeters:sc18434:04}
{Peeters}, E., {Tielens}, A.~G.~G.~M., {Boogert}, A.~C.~A., {Hayward}, T.~L.,
  \& {Allamandola}, L.~J. 2004{\natexlab{c}}, \apj, submitted

\bibitem[{{Peeters} {et~al.}(1999){Peeters}, {Tielens}, {Roelfsema}, \&
  {Cox}}]{Peeters:parijs:99}
{Peeters}, E., {Tielens}, A.~G.~G.~M., {Roelfsema}, P.~R., \& {Cox}, P. 1999,
  The Universe as Seen by ISO, (eds.) Cox P. \& Kessler M.~F., ESA-SP427, 739

\bibitem[{{Petrie} {et~al.}(2003){Petrie}, {Stranger}, \& {Duley}}]{Petrie:03}
{Petrie}, S., {Stranger}, R., \& {Duley}, W.~W. 2003, \apj, 594, 869

\bibitem[{{Piest} {et~al.}(2001){Piest}, {Oomens}, {Bakker}, {von Helden}, \&
  {Meijer}}]{Piest:01}
{Piest}, J.~A.~H., {Oomens}, J., {Bakker}, J., {von Helden}, G., \& {Meijer},
  G. 2001, Spectrochimica Acta, 57, 717

\bibitem[{{Puget} \& {L\'{e}ger}(1989)}]{Puget:revpah:89}
{Puget}, J.~L. \& {L\'{e}ger}, A. 1989, \araa, 27, 161

\bibitem[{{Reach} {et~al.}(2000){Reach}, {Boulanger}, {Contursi}, \&
  {Lequeux}}]{Reach:pahsinsmc:00}
{Reach}, W.~T., {Boulanger}, F., {Contursi}, A., \& {Lequeux}, J. 2000, \aap,
  361, 895

\bibitem[{{Rigopoulou} {et~al.}(1999){Rigopoulou}, {Spoon}, {Genzel}, {Lutz},
  {Moorwood}, \& {Tran}}]{Rigopoulou:99}
{Rigopoulou}, D., {Spoon}, H.~W.~W., {Genzel}, R., {et~al.} 1999, \aj, 118,
  2625

\bibitem[{{Robinson} {et~al.}(1997){Robinson}, {Beegle}, \&
  {Wdowiak}}]{Robinson:97}
{Robinson}, M.~S., {Beegle}, L.~W., \& {Wdowiak}, T.~J. 1997, \apj, 474, 474

\bibitem[{{Roelfsema} {et~al.}(1996){Roelfsema}, {Cox}, {Tielens},
  {Allamandola}, {Baluteau}, {Barlow}, {Beintema}, {Boxhoorn}, {Cassinelli},
  {Caux}, {Churchwell}, {Clegg}, {de Graauw}, {Heras}, {Huygen}, {van der
  Hucht}, {Hudgins}, {Kessler}, {Lim}, \& {Sandford}}]{Roelfsema:pahs:96}
{Roelfsema}, P.~R., {Cox}, P., {Tielens}, A.~G.~G.~M., {et~al.} 1996, \aap,
  315, L289

\bibitem[{{Sakata} {et~al.}(1984){Sakata}, {Wada}, {Tanabe}, \&
  {Onaka}}]{Sakata:QCC:84}
{Sakata}, A., {Wada}, S., {Tanabe}, T., \& {Onaka}, T. 1984, \apjl, 287, L51

\bibitem[{{Schutte} {et~al.}(1993){Schutte}, {Tielens}, \&
  {Allamandola}}]{Schutte:model:93}
{Schutte}, W.~A., {Tielens}, A.~G.~G.~M., \& {Allamandola}, L.~J. 1993, \apj,
  415, 397

\bibitem[{{Sellgren}(1984)}]{Sellgren:84}
{Sellgren}, K. 1984, \apj, 277, 623

\bibitem[{{Sloan} {et~al.}(1997){Sloan}, {Bregman}, {Geballe}, {Allamandola},
  \& {Woodward}}]{Sloan:97}
{Sloan}, G.~C., {Bregman}, J.~D., {Geballe}, T.~R., {Allamandola}, L.~J., \&
  {Woodward}, C.~E. 1997, \apj, 474, 735

\bibitem[{{Snow} \& {Witt}(1995)}]{Snow:c:95}
{Snow}, T.~P. \& {Witt}, A.~N. 1995, Science, 270, 1455

\bibitem[{{Song} {et~al.}(2003{\natexlab{a}}){Song}, {Kerr}, {McCombie},
  {Couch}, \& {Sarre}}]{Song:33:03}
{Song}, I., {Kerr}, T., {McCombie}, J., {Couch}, P., \& {Sarre}, P.
  2003{\natexlab{a}}, \mnras, accepted

\bibitem[{{Song} {et~al.}(2003{\natexlab{b}}){Song}, {McCombie}, {Kerr},
  {Couch}, \& {Sarre}}]{Song:colorado:03}
{Song}, I., {McCombie}, J., {Kerr}, T., {Couch}, P., \& {Sarre}, P.
  2003{\natexlab{b}}, in Astrophysics of Dust, (ed.) Witt A.~N.

\bibitem[{{Tielens}(1990)}]{Tielens:stardust:90}
{Tielens}, A.~G.~G.~M. 1990, in Carbon in the Galaxy: Studies from Earth and
  Space, 59

\bibitem[{{Tielens}(1993)}]{Tielens:93}
{Tielens}, A.~G.~G.~M. 1993, in Dust and Chemistry in Astronomy, (eds.) T.~J.
  {Millar} \& D.~A. {Williams}, 103

\bibitem[{{Tielens}(1997)}]{Tielens:carbonstardust:97}
{Tielens}, A.~G.~G.~M. 1997, \apss, 251, 1

\bibitem[{{Tielens} \& {Allamandola}(1987)}]{Tielens:dust:87}
{Tielens}, A.~G.~G.~M. \& {Allamandola}, L.~J. 1987, in ASSL Vol. 134:
  Interstellar Processes, 397--469

\bibitem[{{Tielens} \& {Charnley}(1997)}]{Tielens:organicmol:97}
{Tielens}, A.~G.~G.~M. \& {Charnley}, S.~B. 1997, Origins Life Evolution
  Biosphere, 27, 23

\bibitem[{{Tielens} {et~al.}(1999){Tielens}, {Hony}, {Van Kerckhoven}, \&
  {Peeters}}]{Tielens:parijs:99}
{Tielens}, A.~G.~G.~M., {Hony}, S., {Van Kerckhoven}, C., \& {Peeters}, E.
  1999, in The Universe as Seen by ISO, (eds.) Cox P. \& Kessler M.~F., ESA-SP427, 579

\bibitem[{{Tokunaga} {et~al.}(1991){Tokunaga}, {Sellgren}, {Smith}, {Nagata},
  {Sakata}, \& {Nakada}}]{Tokunaga:33prof:91}
{Tokunaga}, A.~T., {Sellgren}, K., {Smith}, R.~G., {et~al.} 1991, \apj, 380,
  452

\bibitem[{{Tokunaga} \& {Wada}(1997)}]{Tokunago:97}
{Tokunaga}, A.~T. \& {Wada}, S. 1997, Advances in Space Research, 19, 1009

\bibitem[{{Tran} {et~al.}(2001){Tran}, {Lutz}, {Genzel}, {Rigopoulou}, {Spoon},
  {Sturm}, {Gerin}, {Hines}, {Moorwood}, {Sanders}, {Scoville}, {Taniguchi}, \&
  {Ward}}]{Tran:01}
{Tran}, Q.~D., {Lutz}, D., {Genzel}, R., {et~al.} 2001, \apj, 552, 527

\bibitem[{{Uchida} {et~al.}(2000){Uchida}, {Sellgren}, {Werner}, \&
  {Houdashelt}}]{Uchida:RN:00}
{Uchida}, K.~I., {Sellgren}, K., {Werner}, M.~W., \& {Houdashelt}, M.~L. 2000,
  \apj, 530, 817

\bibitem[{{van Diedenhoven} {et~al.}(2004){van Diedenhoven}, {Peeters}, {Van
  Kerckhoven}, {Hony}, {Allamandola}, {Hudgins}, \&
  {Tielens}}]{vanDiedenhoven:chvscc:03}
{van Diedenhoven}, B., {Peeters}, E., {Van Kerckhoven}, C., {et~al.} 2004,
  \apj, submitted

\bibitem[{{Van Kerckhoven} {et~al.}(2000){Van Kerckhoven}, {Hony}, {Peeters},
  {Tielens}, {Allamandola}, {Hudgins}, {Cox}, {Roelfsema}, {Voors}, {Waelkens},
  {Waters}, \& {Wesselius}}]{VanKerckhoven:plat:00}
{Van Kerckhoven}, C., {Hony}, S., {Peeters}, E., {et~al.} 2000, \aap, 357, 1013

\bibitem[{{Vermeij} {et~al.}(2002){Vermeij}, {Peeters}, {Tielens}, \& {van der
  Hulst}}]{Vermeij:pahs:01}
{Vermeij}, R., {Peeters}, E., {Tielens}, A.~G.~G.~M., \& {van der Hulst}, J.~M.
  2002, \aap, 382, 1042

\bibitem[{{Verstraete} {et~al.}(1990){Verstraete}, {Leger}, {D'Hendecourt},
  {Defourneau}, \& {Dutuit}}]{Verstraete:90}
{Verstraete}, L., {Leger}, A., {D'Hendecourt}, L., {Defourneau}, D., \&
  {Dutuit}, O. 1990, \aap, 237, 436

\bibitem[{{Verstraete} {et~al.}(2001){Verstraete}, {Pech}, {Moutou},
  {Sellgren}, {Wright}, {Giard}, {L{\' e}ger}, {Timmermann}, \&
  {Drapatz}}]{Verstraete:prof:01}
{Verstraete}, L., {Pech}, C., {Moutou}, C., {et~al.} 2001, \aap, 372, 981

\bibitem[{{Verstraete} {et~al.}(1996){Verstraete}, {Puget}, {Falgarone},
  {Drapatz}, {Wright}, \& {Timmermann}}]{Verstraete:m17:96}
{Verstraete}, L., {Puget}, J.~L., {Falgarone}, E., {et~al.} 1996, \aap, 315,
  L337

\bibitem[{{Wada} {et~al.}(2003){Wada}, {Onaka}, {Yamamura}, {Murata}, \&
  {Tokunaga}}]{Wada:13c:03}
{Wada}, S., {Onaka}, T., {Yamamura}, I., {Murata}, Y., \& {Tokunaga}, A.~T.
  2003, \aap, 407, 551

\bibitem[{{Wagner} {et~al.}(2000){Wagner}, {Kim}, \& {Saykally}}]{Wagner:2000}
{Wagner}, D.~R., {Kim}, H., \& {Saykally}, R.~J. 2000, \apj, 545, 854

\bibitem[{{Witteborn} {et~al.}(1989){Witteborn}, {Sandford}, {Bregman},
  {Allamandola}, {Cohen}, {Wooden}, \& {Graps}}]{Witteborn:89}
{Witteborn}, F.~C., {Sandford}, S.~A., {Bregman}, J.~D., {et~al.} 1989, \apj,
  341, 270

\bibitem[{{Young Owl} {et~al.}(2002){Young Owl}, {Meixner}, {Fong}, {Haas},
  {Rudolph}, \& {Tielens}}]{YoungOwl:02}
{Young Owl}, R.~C., {Meixner}, M.~M., {Fong}, D., {et~al.} 2002, \apj, 578, 885

\end{thebibliography}
\end{document}